\documentclass{article}

\usepackage{microtype}
\usepackage{graphicx}
\usepackage{subfigure}
\usepackage{booktabs} 

\usepackage{hyperref}

\usepackage[accepted]{icml2023}

\usepackage{amsmath}
\usepackage{amssymb}
\usepackage{mathtools}
\usepackage{amsthm}

\usepackage[capitalize,noabbrev]{cleveref}

\theoremstyle{plain}
\newtheorem{theorem}{Theorem}[section]

\newtheorem{lemma}[theorem]{Lemma}
\newtheorem{corollary}[theorem]{Corollary}
\theoremstyle{definition}
\newtheorem{definition}[theorem]{Definition}

\theoremstyle{remark}

\newtheorem{claim}{Claim}
\newtheorem{observation}{Observation}

\usepackage{color}

\newcommand{\card}[1]{|#1|}

\newcommand{\E}{\mathbf{E}}
\newcommand{\R}{\mathbb{R}}
\newcommand{\Z}{\mathbb{Z}}

\newcommand{\stt}{\textnormal{s.t.}}
\newcommand{\A}{\mathcal{A}}
\newcommand{\B}{\mathcal{B}}
\newcommand{\D}{\mathcal{D}}
\newcommand{\F}{\mathcal{F}}
\newcommand{\X}{\mathcal{X}}
\newcommand{\Y}{\mathcal{Y}}
\renewcommand{\S}{\mathcal{S}}
\newcommand{\C}{\mathcal{C}}

\newcommand{\cost}{\textbf{cost}}
\newcommand{\dom}{\textbf{dom}}

\newcommand{\opt}{\text{OPT}}
\newcommand{\alg}{\text{ALG}}
\newcommand{\dep}{\text{depth}}

 \newenvironment{opf}
{\begin{trivlist} \item[] {\em Proof of Observation. }}
	{$\hfill\diamond$ \end{trivlist}}

\newenvironment{cpf}
{\begin{trivlist} \item[] {\em Proof of Claim. }}
	{$\hfill\diamond$ \end{trivlist}}

\usepackage[textsize=tiny]{todonotes}

\icmltitlerunning{Buying Information for Stochastic Optimization}

\begin{document}

\twocolumn[
\icmltitle{Buying Information for Stochastic Optimization}




\begin{icmlauthorlist}
\icmlauthor{Mingchen Ma}{yyy}
\icmlauthor{Christos Tzamos}{yyy}
\end{icmlauthorlist}

\icmlaffiliation{yyy}{Department of Computer Sciences, University of Wisconsin-Madison, Madison, WI, USA}

\icmlcorrespondingauthor{Mingchen Ma}{mma54@wisc.edu}

\icmlkeywords{Machine Learning, ICML}

\vskip 0.3in
]



\printAffiliationsAndNotice{\icmlEqualContribution} 

\begin{abstract}

Stochastic optimization is one of the central problems in Machine Learning and Theoretical Computer Science. In the standard model, the algorithm is given a fixed distribution known in advance. In practice though, one may acquire at a cost extra information to make better decisions.
In this paper, we study how to buy information for stochastic optimization and formulate this question as an online learning problem. Assuming the learner has an oracle for the original optimization problem, we design a $2$-competitive deterministic algorithm and a $e/(e-1)$-competitive randomized algorithm for buying information. We show that this ratio is tight as
    the problem is equivalent to a robust generalization of the ski-rental problem, which we call super-martingale stopping.  
 We also consider an adaptive setting where the learner can choose to buy information after taking some actions for the underlying optimization problem. We focus on the classic optimization problem, Min-Sum Set Cover, where the goal is to quickly find an action that covers a given request drawn from a known distribution. We provide an $8$-competitive algorithm running in polynomial time that chooses actions and decides when to buy information about the underlying request.

\end{abstract}

\section{Introduction}

\subsection{Offline and Adaptive Stochastic Optimization}

Stochastic optimization is one of the core problems in machine learning and theoretical computer sciences. In stochastic optimization, the input parameters of the problems are random variables drawn from a known distribution. Given the distribution of the parameters, a learner constructs a feasible solution in advance (offline stochastic optimization) or adaptively (adaptive stochastic optimization) to optimize the objective function in expectation. Formally, the two types of stochastic optimization problems can be defined in the following way.

\begin{definition}[Offline Stochastic Optimization]
    Let $\S$ be a set of scenarios and $\X(\S)$ be a set of actions. Let $\ell(A,s): 2^\X \times \S \to R_+$ be a loss function.  An offline stochastic optimization problem $(\X,\S,\ell,\D)$ is to find a set of actions $A$ that minimize $\E_{s \sim \D}\ell(A,s)$, where $\D$ is a distribution over $\S$.
\end{definition}

\begin{definition}[Adaptive Stochastic Optimization]
    Let $\S$ be a set of scenarios and $\X(\S)$ be a set of actions. 
    Initially, a random scenario $s$ is drawn according to a distribution $\D$.
    Then, the learner sequentially chooses actions $a_1,a_2,\dots$ and after the $t$-th action  observes a (possibly randomized) outcome $r( (a_1,a_2,\dots,a_t), s) \in \R$. The goal of the learner is to take a sequence of actions $A$ that minimizes $\E_{s \sim \D} \ell(A,s)$ for a given loss function $\ell(A,s)$, possibly exploiting the information gained about $s$ along the way.
    
    
    
    \end{definition}

A huge body of work among different communities such as machine learning, theoretical computer science, statistics, and operations research has studied stochastic optimization problems given their numerous applications.
For example, methods of offline stochastic optimization have been widely applied to problems such as training machine learning models \cite{shalev2009stochastic,bottou2010large,kingma2014adam} and mechanism design \cite{nisan1999algorithmic,hartline2013mechanism,roughgarden2016twenty}. On the other hand, many adaptive stochastic optimization problems such as Pandora's Box problem \cite{weitzman1979optimal,chawla2020pandora}, active learning \cite{dasgupta2004analysis,settles2012active} and optimal decision tree \cite{adler2012approximating,li2020tight} have also been applied to areas like artificial intelligence, microeconomics, and operations research.  

A common assumption in these works is that the distribution $\D$ of the scenario $s$ is considered as a given. However, such an assumption is not realistic in practice. A learner in practice has many ways to gain extra knowledge on the optimization problem he is going to solve. With the extra knowledge, it is reasonable that the learner updates the prior distribution $\D$ to some posterior distribution $\D'$ and uses a better strategy to solve the problem. 
As a concrete example, consider $n$ bidders that compete over an item in an auction. 
Classic auction theory assumes that the auctioneer only knows a prior distribution $\D$ over the buyer values and wants to design an auction to optimize a target objective such as welfare or revenue. In practice though, there is a number of information sources available to the auctioneer that provide information about the bidders such as their demographics, their preferences or their purchase history. Such information can be very useful.


However, this information does not come for free. It may cost significant amounts of money or time and it is not clear in advance, how helpful this information will be. In the example, the auctioneer may pay an information provider only to receive irrelevant pieces of information or information already known. 
\subsection{Our Contribution and Techniques}

In this paper, we study the problem of buying information for stochastic optimization. We consider a learner that wants to minimize the total cost spent on solving the optimization problem and the cost of acquiring information.

We model the information acquisition process using a signaling scheme \citep{emek2014signaling}. A signaling scheme is a (randomized) function $f$ from the set of scenarios $\S$ to a signal space $\R$. If a learner asks for feedback from $f$, he will receive a signal $y$ and the prior distribution can be updated as $\D|_{f(s)=y}$. In our model, we assume there is a sequence of signaling schemes $\F=\{f_t\}_{t=0}^\infty$ arriving online. 
At any timestep $t$, based on the signals received so far, the learner has the choice to continue purchasing the next signal given by $f_t$ or stop.
Our goal is to construct a learner who is competitive to the cost of a prophet who knows the structure of $\F$ in advance and can take optimal actions.


For offline optimization, all signals must be purchased before taking any actions in the underlying stochastic optimization problem. 
We assume the learner is able to compute an (approximate) optimal solution for the underlying problem given the available information at any point in time. The goal of the learner is then to adaptively decide when to stop buying feedback. Our main results in this setting are summarized below:

\begin{theorem}[Informal Version of Theorem~\ref{th martingale det} and Theorem~\ref{th martingale rand}]
There exist a $2$-competitive deterministic learner and an $\frac{e}{e-1}$-competitive randomized learner to buy information for offline stochastic optimizations.
\end{theorem}

We show that both learners can be implemented efficiently and have competitive ratios that are information theoretically optimal. Thus, we give a comprehensive understanding of buying feedback for offline stochastic optimization. 
To solve the problem, we formulate it as a super-martingale stopping problem:
There is an unknown sequence of random variables $(X_0, X_1,\dots)$ satisfying $\E(X_{i+1} \mid X_i) \le X_i$. The realizations of the random variables arrive online and an algorithm outputs a stopping index $i^*$ adaptively to minimize $\E (i^*+X_{i^*})$. The super-martingale stopping problem can be seen as a generalization of the classic ski-rental problem introduced in \cite{karlin1994competitive} where all $X_i \in \{0,B\}$ and its variant introduced in \cite{chawla2020pandora}, where $(X_0,X_1,\dots)$ are monotone decreasing constants. In the more general setting of super-martingale stopping though, the values of $X_i$ may not be monotone, and they are only monotone in expectation. This makes the problem significantly more challenging and as we show in Appendix~\ref{sec counter}, natural algorithms for ski-rental problems are not competitive for our problem. 

For adaptive stochastic optimization, 
it is also natural to intertwine  purchasing information with taking actions. For example, several actions may be taken first in the problem and then information may be purchased conditional on their outcome. As this setting is more problem dependent, we focus on a paradigmatic case of adaptive stochastic optimization, where there is a random set of good actions, and the learner takes actions in each round until a good action is chosen. Such a problem is called Min Sum Set Cover (MSSC), a well-studied adaptive stochastic optimization problem \citep{bar1998chromatic,bar1999matched,feige2004approximating}.
In our model, the learner has an extra action at each round to buy information getting a better estimate of the probability that an action is good. 

We provide an algorithm for this problem competitive to a prophet that knows the sequence of signaling schemes in advance:
\begin{theorem}[Informal Version of Theorem~\ref{th costly up}]
    There is a poly-time learner that is $8$-competitive for buying information for Min Sum Set Cover.
\end{theorem}
 We achieve this in two steps. In the first step, we show we can shrink the action space so that we don't need to consider when to buy feedback. We introduce a simpler model called adaptive stochastic optimization with time dependent feedback, where a learner takes an action in each round, and feedback arrives for free after an action is taken. We show in Theorem~\ref{th reduction} that if there is a learner that is $\alpha$-competitive for adaptive stochastic optimization with time dependent feedback, then we can use it to construct a $2\alpha$-competitive learner to buy information for adaptive stochastic optimization. Our second step is to prove the following technical theorem, which is of independent interest.
 \begin{theorem}[Informal Version of Theorem~\ref{th time dependent up}]
     The greedy algorithm is $4$-competitive for MSSC with time dependent feedback.
 \end{theorem}
There is a lot of work done for analysis of the greedy algorithm of min sum coverage objective under different settings \citep{feige2004approximating,streeter2008online,golovin2011adaptive}. The analysis is usually based on an elegant histogram approach proposed in \cite{feige2004approximating}. However, in our model, the decision made by the learner is fully adaptive and it is hard to adapt such an analysis directly. Instead, we bypass such difficulty and use an interesting linear programming dual approach to analyze the greedy algorithm.
Besides algorithmic results, we also present hard instances to build information theoretic lower bound for MSSC under our models.

\subsection{Applications of our Model}

Buying information is very common in practice. In fact, our model fits well in both theory and practical applications. In this section, we give several applications of our model.  We first give a typical example of buying information for offline stochastic optimization.
\paragraph{Selling One Item with Feedback}
    There is a seller who wants to sell an item to a buyer. The seller sets a price $p$ for the item. The buyer has a value $v$ for the item and would like to pay the price $p$ for the item if $p \le v$. However, if $p>v$, the buyer will not buy the item. Given a pair of $(v,p)$, denote by $P(v,p)=p \mathbf{1}_{p \le v}$ the payment of the buyer. The value of the buyer may depend on his nationality, education, or other factors. The information can be collected from the historic trade and thus the seller has a prior distribution $\D$ of the value $v$. The goal of the learner is to set up the price $p$ to minimize $\E_{v \sim \D}\left(v-P(v,p)\right)$. However, instead of setting the price immediately, the seller may pay some money to collect more information about the buyer. This can help the seller update the prior distribution of the value $v$. In practice, it is hard to predict the quality of the information. The question for the seller is how much information is sufficient for him to set up a good price.

Our second example is on buying information for adaptive stochastic optimization.

\paragraph{Optimal Decision Tree with Feedback}
A doctor wants to diagnose the disease of a patient. There are $\X=[n]$ different tests that can be performed by the doctor and $\S=[m]$ different possible diseases. If the patient has a disease $s \in \S$ and a test $a\in \X$ is performed, then the doctor will receive an outcome $r(s,a)$. The doctor has a prior distribution $\D$ of the disease $s$ based on the symptom of the patient. In the standard optimal decision tree problem, based on the knowledge of $\D$, the goal of the doctor is to perform a sequence of tests adaptively to identify the disease while minimizing the expected cost of the tests. 
In practice, the doctor may choose not to run tests but instead send the patient home to see whether the symptoms worsen.
However, this is also costly and it may be challenging to predict what symptoms will appear and how much time it will take for them to appear. Combined with an algorithm for computing approximately optimal decision trees, our work shows how to incorporate the symptom monitoring component to efficiently identify the disease.

Beyond these applications, our model  fits well with many existing theoretical frameworks in learning theory. Here we take adaptive submodular optimization, a recently popular research direction in the field of machine learning as our example. 

\paragraph{Adaptive Submodularity with Feedback}
Motivated by applications on artificial intelligence, \cite{golovin2011adaptive} introduces the notion of adaptive submodularity, which was a popular research topic in the last decade. A function $f(A,s)$ of a set of actions $A$ and a random scenario $s$ is adaptive submodular if $\E_s f(A,s)$ is a submodular function. After an action $a$ is taken, the learner will see an outcome $s(a)$. Given the distribution of $s$, the learner will construct the action set $A$ adaptively to optimize classic objectives for submodular functions \citep{fujishige2005submodular} such as submodular maximization, min submodular coverage, and min sum submodular coverage. Many natural questions arise when feedback is involved in this framework. For example, if feedback is costly, how can we buy feedback to help us make adaptive decisions? If the feedback is free and time dependent, are existing policies still competitive?

\subsection{Organization of paper}
In Section~\ref{sec model}, we formally introduce the model studied by the paper. In Section~\ref{sec martingale}, we introduce the super-martingale stopping problem to study buying information for offline stochastic optimization. We give a tight deterministic algorithm and a tight randomized algorithm for the super-martingale stopping problem. Furthermore, we will discuss the robustness of these algorithms. In Section~\ref{sec reduction}, we focus on buying information for adaptive stochastic optimization.
We introduce the model of time dependent feedback and build a connection between adaptive stochastic optimization with time dependent feedback and buying information for adaptive stochastic optimization in Section~\ref{sec time dependent}. In Section~\ref{sec mssc}, we show a simple greedy learner is $4$-competitive for Min Sum Set Cover with time dependent feedback. And in Section~\ref{sec mssc cost}, we design an $8$-competitive algorithm for buying information for Min Sum Set Cover. Furthermore, we discuss the information theoretic lower bound for Min Sum Set Cover under both settings.

\section{Stochastic Optimization with Feedback}\label{sec model}

\subsection{Feedback Signals for Stochastic Optimization} \label{sec feedback}

 Let $\S$ be a set of scenarios with a distribution $\D$ over $\S$ and let $\Y$ be a set of random variables over $\R$. A randomized signaling scheme $f: \S \to \Y$ is a map from $\S$ to $\Y$. Let $s$ be a scenario drawn from $\D$. A signal received from $f$ is a realization $y \in \R$ of the random variable $f(s)$. 
Similarly, a deterministic signaling scheme $f: \S \to \R$ is a function from $\S$ to $\R$. When a scenario $s$ is drawn, a signal received from $f$ is defined by $y=f(s) \in \R$. In particular, any deterministic signaling scheme gives a partition of $\S$. Given the definition of a signaling scheme, we are able to define feedback for stochastic optimization problems.

\begin{definition}[Feedback]
Let $(\X,\S,\ell,\D)$ be a stochastic optimization problem. A sequence of feedback $\F=\{f_t\}_{t=0}^\infty$ is a sequence of \textbf{unknown} randomized (deterministic) signaling scheme. The $t$th feedback received by a learner is the pair $(y_{t},\D\mid_{f_t(s)=y_t, f_{t-1}(s)=y_{y-1},\dots,f_0(s)=y_0})$, where $y_t$ is the signal from $f_t$.   
\end{definition}
For convenience, we assume $f_0$ is a constant for every scenario, throughout the paper. Such an assumption is used to reflect the fact that the learner has no extra knowledge at time 0.
In fact, for our model, randomized signaling schemes are equivalent to deterministic ones. We leave a discussion for this in Appendix~\ref{sec signal}.
In this paper, we consider deterministic signaling schemes. A deterministic signaling scheme can simplify our analysis and provide more intuition.
In particular, if each signaling scheme $f \in \F$ is deterministic, then $\F$ can be represented as a tree. For such feedback $\F$, we define a feedback tree $T(\F)$ as follows.

\begin{definition}[Feedback Tree]
    Let feedback $\F=\{f_t\}_{t=0}^\infty$ be a set of deterministic signaling schemes. The feedback tree $T(\F)$ for $\F$ is a tree that is defined as follows. Each node $v \in T(\F)$ contains a set of scenarios and the children of $v$ form a partition of the set of scenarios contained in $v$. The root of $T(\F)$ contains all scenarios. For every $s \in \S$, let $P(s)=(v_0,v_1,\dots,v_n)$ be the longest path in $T(\F)$ such that every node in $P(s)$ contains $s$. Then the set of scenarios contained in $v_i$ is defined by $\{s' \in v_{i-1} \mid f_i(s')=f_i(s)\}$.
\end{definition}

\subsection{Problem Formulation}

Although feedback is helpful for a learner to make better decisions for stochastic optimization problems, obtaining feedback always requires some cost. The cost can be either time or money. Thus, it is natural for a learner to consider how to balance the cost of asking for feedback and the cost of solving the optimization problem. We consider formulating this problem in an online fashion for offline and adaptive stochastic optimization problems.


\begin{definition}[Buying Information for Offline Stochastic Optimization]
    Let $(\X,\S_0,\ell,\D_0)$ be an offline stochastic optimization problem and $\F=\{f_t\}_{t=0}^\infty$ be a sequence of \textbf{unknown} feedback. Let $\C=\{c_t\}_{t=0}^\infty$ be a sequence of cost for receiving a signal from $f_{t+1} \in \F$. Here, $c_t: \R \to \Z_+$ is a nonnegative function that depends on the last received signal. In each time round $t \ge 0$, a learner receives an offline stochastic optimization problem $(\X,\S_t,\ell,\D_t)$ and a cost $c_t(y_t)$ to obtain a signal from $f_{t+1}$, where $y_t$ is the signal received from $f_t$. Here, $\S_t=\{s \in \S_{t-1} \mid y_t \in \dom( f_{t}(s))\}$ and $\D_t=\D_{t-1}\mid_{f_t(s)=y_t}$ for $t \ge 1$. The learner can either stop and pay $\sum_{j=0}^{t-1}c_j(y_{j})+\min_{A \subseteq \X} \E_{s \sim \D_t}\ell(A,s)$ or enter the next time round. An offline stochastic optimization with feedback $(\X,\S,\ell,\D,\F,\C)$ is to decide a stopping time $T$ adaptively to minimize $\E_T \left(  \sum_{j=0}^{T-1}c_j(y_{j})+\min_{A \subseteq \X} \E_{s \sim \D_T}\ell(A,s) \right)$.
\end{definition}

Let $I=(\X,\S,\ell,\D,\F,\C)$ be an instance of offline stochastic optimization with feedback, denote by $\cost(\A,I)$ the cost of the stopping time output by a learner $\A$ for the given instance. 
A learner is $\alpha$-competitive if for every instance $(\X,\S,\ell,\D,\F,\C)$, $\cost(\A,I) \le \alpha \opt(I) = \alpha \min_\A \cost(\A,I)$.

We can describe the problem in a more intuitive way in terms of the feedback tree. Let $(\X,\S,\ell,\D)$ be a stochastic optimization problem and $T(\F)$ be a feedback tree. Each node $v$ of $T(\F)$ represents a new stochastic optimization problem $(\X,\S_v,\ell,\D_v)$, where $\S_v$ is the set of scenarios contained in $v$ and $\D_v=\D\mid_{s \in \S_v}$. Solving this optimization problem needs a cost $\min_{A \subseteq \X} \E_{s \sim \D_v}\ell(A,s)$.
Each node also has a cost $c_v$ to move down for one step. The stochastic optimization problem and the cost will be revealed to the learner when the learner reaches $v$. $T(\F)$ is unknown to the learner and a path of $T(\F)$ is selected according to $\D$ initially. The learner will keep moving along the path by paying the cost $c_v$ and will decide when to stop and solve the optimization problem. The benchmark we want to compare is a learner who knows the whole feedback tree in advance and thus can compute the optimal stopping time.


\begin{definition}[Buying Information for Adaptive Stochastic Optimization]
    Let $(\X,\S,\ell,\D)$ be an adaptive stochastic optimization problem. $\F=\{f_t\}_{t=0}^\infty$ be a sequence of \textbf{unknown}   feedback. Let $\C=\{c_t\}_{t=0}^\infty$ be a sequence of cost for receiving a signal from $f_{t+1} \in \F$. Here, $c_t: \R \to \Z_+$ is a nonnegative function that depends on the last received signal.
     Initially, a scenario $s$ is drawn according to $\D$. In each time round $t$, a learner first adaptively receives an arbitrary number of signals $y(s)$ from the sequence $\F$ by paying the corresponding cost, then selects an action $a_t \in \X$. Let $T(s)$ be the number of signals received by the learner if $s$ is drawn. An adaptive stochastic optimization problem with feedback is to make decisions to ask for feedback and take actions adaptively in each time round to minimize $\E_{s \sim \D} \left(\ell(A,s)+\sum_{j=0}^{T(s)-1}c_j(y_j(s))\right)$.
\end{definition}
Let $I=(\X,\S,\ell,\D,\F,\C)$ be an instance of adaptive stochastic optimization with feedback, denote by $\cost(\A,I)$ the expected cost of the decisions made by a learner $\A$ for the given instance. 
A learner is $\alpha$-competitive if for every instance $(\X,\S,\ell,\D,\F,\C)$, $\cost(\A,I) \le \alpha \opt(I) = \alpha \min_\A \cost(\A,I)$.

\section{Buying Information for Offline Stochastic Optimization and Super-Martingale Stopping Problem}\label{sec martingale}



Let $(\X,\S,\ell,\D)$ be an offline stochastic optimization problem and $f$ be a signaling scheme. Denote by $\D_y$ the posterior distribution of $\D$ after receiving signal $y$ from $f$. Although it is possible that $\min_{A\subseteq \X} \E_{s\sim \D}\ell(A,s)< \min_{A \subseteq \X}\E_{s \sim \D_y}\ell(A,s)$, it is always true that 
\begin{align*}
    \E_y\min_{A\subseteq \X} \E_{s\sim \D_y}\ell(A,s) \le  \min_{A \subseteq \X}\E_{s \sim \D}\ell(A,s).
\end{align*}
That is to say, feedback is always helpful in expectation. This implies the sequence of minimum value of the stochastic optimization problems is a super-martingale. Formally,
given a sequence of feedback $\F$, denote by $D_i$ the posterior distribution after receiving signals from $f_0,f_1,\dots,f_i$. Let random variable $X_i = \min_{A \subseteq \X}\E_{s \sim \D_i}\ell(A,s)$. Then for every $i \ge 0$, we have $\E\left(X_{i+1} \mid X_i\right) \le X_i$. This motivates us to formulate the problem of buying information as the following super-martingale stopping problem. As we discuss in Appendix~\ref{sec formulation}, super-martingale stopping  is equivalent to buying information for stochastic optimization.

\subsection{Super-Martingale Stopping Problem}

\begin{definition}[Super-Martingale Stopping Problem]
    Let $X_0,X_1,\dots,X_n$ be a sequence of nonnegative random variables unknown to the learner. Assume for every $i$, $\E(X_{i+1} \mid X_i) \le X_i$. The problem has $n+1$ rounds. In the $i$th round, given an observed realization of $X_0,\dots,X_i$, a learner decides either to stop and pay $i+X_i$ or to obtain the realization of $X_{i+1}$ and go to the next round. The goal of the learner is to compute a decision rule to obtain a stopping time $i^*$ \textbf{only based on the observed realization of the sequence}  to minimize $\E(i^*+X_{i^*})$.
\end{definition}
For convenience, we assume $X_0$ is a constant throughout the paper.
Suppose each random variable $X_i$ has finite support, then the sequence can be represented by a tree $T$, where a node $v$ with depth $i$ stores a realization of $X_i$. To simplify the notation, we use $v$ to denote both the node and the value stored at the node.
When we make a single movement from node $v$, we will reach a child $v'$ of $v$ with probability $\Pr(X_{i+1}=v'\mid X_i=v)$. An optimal learner knows tree $T$ in advance and can decide in advance which node to stop to optimize the expected cost. Formally, a set of stopping nodes $S$ is feasible for $T$ if every path of $T$ with length $n$ contains one and only one stopping node. The cost of $S$ is $\sum_{v \in S}\Pr(v)(\dep (v)+v)$. We denote by $\opt(T)$ the minimum cost among all feasible sets of stopping nodes of $T$. An algorithm is $\alpha$-competitive if for every instance of the super-martingale stopping problem with a representation $T$, the expected cost of the algorithm $ \alg(T)=\E(i^*+X_{i^*}) \le \alpha \opt(T)$.

In the ski-rental problem studied in \cite{karlin1994competitive}, there is a pair of positive numbers $(B,T)$ such that $X_i=B$ if $i<T$ and $X_i=0$ if $i \ge T$.
This implies that ski-rental problem is a special case of the super-martingale stopping problem. Thus, we have the following information theoretic lower bound for the super-martingale stopping problem.

\begin{theorem}\label{th lb rand}
    For every $\epsilon>0$, no randomized algorithm is $\frac{e}{e-1}-\epsilon$-competitive for the super-martingale stopping problem.
\end{theorem}

\begin{theorem}\label{th lb det}
        For every $\epsilon>0$, no deterministic algorithm is $2-\epsilon$-competitive for super-martingale stopping problem.
\end{theorem}

Recall that the key idea in the design of algorithms for ski-rental problem is to balance the payment $X_i$ and the index $i$. However, this idea cannot be simply applied to the super-martingale stopping problem.
There are two difficulties faced in the super-martingale stopping problem. First, since any algorithm can only get information from one path of the tree, it is hard to estimate the expected stopping time for the whole tree. Second, unlike most ski-rental type problems, the value $X_i$ is not necessarily decreasing. It is possible that an algorithm moves for one step but sees an $X_i$ with a very large value.  We will show in Appendix~\ref{sec counter} that some natural algorithms that work for ski-rental problems are not competitive for the super-martingale stopping problem. On the other hand, in Appendix~\ref{sec simple}, we establish a simple randomized $2$-competitive algorithm for the super-martingale stopping problem using a completely novel idea. Although the algorithm we present in Appendix~\ref{sec simple} shows competitive algorithms do exist for super-martingale stopping problem, the competitive ratio doesn't match the information theoretic lower bound in Theorem~\ref{th lb rand} and Theorem~\ref{th lb det}.  
In the following sections, we will give a tight deterministic algorithm and randomized algorithm for the super-martingale stopping problem. Furthermore, we will also discuss the robustness of these algorithms, when the input is not a super-martingale.

 The key idea for designing our algorithms is to maintain the following estimator $Q_p(t)$ throughout the execution of the algorithms.
 Let $(X_1,\dots,X_n)$ be an instance of super-martingale stopping and let $T$ be its tree representation.
Initially, a path $p=(v_0,v_1,\dots,v_n)$ of $T$ will be drawn randomly according to the joint distribution of $(X_0, \dots, X_n)$. We define a function $v_p(t)=v_i,$ if $t \in [i,i+1)$. Furthermore, we define $Q_p(t)=\int_0^t\frac{1}{v_p(t)}dt$. In particular, $Q_p(t)$ only depends on our observed realization and doesn't depend on the realization of the random variables we have not seen.  We notice that $Q_p(t)$ is strictly increasing with respect to $t$ and thus for every $s \ge 0$, we can define its inverse function $Q^{-1}_p(s)=t,$ where $Q_p(t)=s$. The power of $Q$ is that it can be used to upper bound and lower bound the optimal stopping time, which can be summarized by the following two lemmas that we will frequently used in our proof. The proof of Lemma~\ref{lm up time} can be found in Appendix~\ref{sec pflm up time} due to a lack of space.

\begin{lemma}\label{lm lb time}

    Let $T$ be a tree representation of an instance of the super-martingale stopping problem and let $p$ be a path of $T$. Then for every $s>r>0$, $Q^{-1}_p(s)-Q^{-1}_p(r) = \int_r^s v_p(Q^{-1}_p(w))dw$.
\end{lemma}

\begin{proof}
    The proof follows a change of variable. We write $w=Q_p(t)$. Then we have 
    \begin{align*}
        & \int_r^s v_p(Q^{-1}_p(w))dw = \int_{Q^{-1}_p(r)}^{Q^{-1}_p(s)} v_p(t)dQ_p(t) \\
        &= \int_{Q^{-1}_p(r)}^{Q^{-1}_p(s)} \frac{v_p(t)}{v_p(t)}dt =  Q^{-1}_p(s)-Q^{-1}_p(r).
    \end{align*}
\end{proof}

\begin{lemma}\label{lm up time}
    Let $v \in T$ be a node with depth $i$ and let $\{p_j\}_{j=1}^k$ be the set of paths that passes $v$. For every $\rho^* \in [Q_{p_j}(i),Q_{p_j}(i+1)]$ and for every $\rho \ge \rho^*$, $\sum_{j=1}^k\Pr(p_j)(Q_{p_j}^{-1}(\rho)-Q_{p_j}^{-1}(\rho^*)) \le \Pr(v)(\rho-\rho^*)v$. 
\end{lemma}

\subsection{A Tight Deterministic Algorithm for Martingale Stopping}

In this section, we propose a simple deterministic 2-competitive algorithm for the super-martingale stopping problem. The competitive ratio is tight according to Theorem~\ref{th lb det}. We leave the proof for Appendix~\ref{sec pf martingale det} due to the space limit.

\begin{theorem}\label{th martingale det}
    There is a deterministic poly-time algorithm that is $2$-competitive for the super-martingale stopping problem.
\end{theorem}

	\begin{algorithm}
		\caption{\textsc{DeterministicStopping} ($2$-competitive deterministic algorithm for super-martingale stopping)}\label{alg detstop}
		\begin{algorithmic}
  \FOR{i=0,1,2,\dots}
  
\STATE Observe $X_i=v_i$ and compute $Q_p(t)$ for $t \in [i,i+1]$ based on the realization of $X_0,\dots,X_i$.
\IF{$\exists t\in (i,i+1]$ such that $Q_p(t)=1$}

\STATE Stop at time $i$ and pay $i+v_{i}$. 

\ENDIF
  
  \ENDFOR
		\end{algorithmic}
	\end{algorithm}

In particular, if the sequence of random variables is monotone decreasing, then our algorithm can even compete against a prophet who knows the realization of the sequence in advance.

\begin{corollary}\label{cor detstop}
Let $I$ be an instance of the super-martingale stopping problem and $(X_0,X_1,\dots)$ be the input sequence. Denote by $\alg(I)$ the cost of Algorithm~\ref{alg detstop} over instance $I$. If $(X_0,X_1,\dots)$ is monotone decreasing, then $\alg(I) \le 2 \E \min_i \left(i+X_i\right)$.
\end{corollary}

\begin{proof}
    Let $x=(x_0,x_1,\dots)$ be a realization of $(X_0,X_1,\dots)$ and denote by $\alg(x)$ the cost of Algorithm~\ref{alg detstop} if the realization is $x$. Since $x$ is monotone decreasing, we have $\alg(x) \le 2\min_i \left(i+x_i\right)$. Thus,
    \begin{align*}
        \alg(I)  \le \E_x 2\min_i \left(i+x_i\right) = 2 \E \min_i \left(i+X_i\right).
    \end{align*}
\end{proof}

\subsection{A Tight Randomized Algorithm for Martingale Stopping}

In this section, we extend the idea of Theorem~\ref{th martingale det} to obtain a $\frac{e}{e-1}$-competitive randomized algorithm for the super-martingale stopping problem. Notice that according to Theorem~\ref{th lb rand}, the competitive ratio is tight. Recall that in the Algorithm~\ref{alg detstop}, we maintain an estimator $Q_P(t)$ throughout the execution of the algorithm and stop when $Q_P(t)=1$. To obtain a better randomized algorithm, we select a random threshold $\rho$ initially, and stop when $Q_P(t)$ exceeds this threshold. The proof of Theorem~\ref{th martingale rand} can be found in Appendix~\ref{sec pf martingale rand}.

\begin{theorem}\label{th martingale rand}
    There is a randomized poly-time algorithm for the super-martingale stopping problem that is $\frac{e}{e-1}$-competitive.
\end{theorem}

	\begin{algorithm}
		\caption{\textsc{RandomizedStopping} ($\frac{e}{e-1}$-competitive algorithm for super-martingale stopping)}\label{alg randstop}
		\begin{algorithmic}
		\STATE Randomly draw a threshold $\rho \in [0,1]$ with a probability density function $p(\rho)=\frac{e^\rho}{e-1}$.
  \FOR{i=0,1,2,\dots}
  
\STATE Observe $X_i=v_i$ and compute $Q_p(t)$ for $t \in [i,i+1]$ based on the realization of $X_0,\dots,X_i$.
\IF{$\exists t\in [i,i+1]$ such that $Q_p(t)=\rho$}

\STATE Stop at time $t$ and pay $i+v_i$. \COMMENT{Every time we stop, $i \le t$.}

\ENDIF
  
  \ENDFOR
		\end{algorithmic}
	\end{algorithm}

Similarly, we have the following corollary, when the input sequence is monotone decreasing.

\begin{corollary}\label{cor randstop}
Let $I$ be an instance of the super-martingale stopping problem and $(X_0,X_1,\dots)$ be the input sequence. Denote by $\alg(I)$ the cost of Algorithm~\ref{alg randstop} over instance $I$. If $(X_0,X_1,\dots)$ is monotone decreasing, then $\alg(I) \le \frac{e}{e-1} \E \min_i \left(i+X_i\right)$.
\end{corollary}

\begin{proof}
    Let $x=(x_0,x_1,\dots)$ be a realization of $(X_0,X_1,\dots)$ and denote by $\alg(x)$ the cost of Algorithm~\ref{alg randstop} if the realization is $x$. Since $x$ is monotone decreasing, we have $\alg(x) \le \frac{e}{e-1}\min_i \left(i+x_i\right)$. Thus,
    \begin{align*}
        \alg(I)  \le \E_x \frac{e}{e-1}\min_i \left(i+x_i\right) = \frac{e}{e-1} \E \min_i \left(i+X_i\right).
    \end{align*}
\end{proof}

\subsection{A Discussion on Benchmark}\label{sec benchmark}
In this section, we discuss the benchmark of the super-martingale stopping problem. According to Corollary~\ref{cor detstop} and Corollary~\ref{cor randstop}, if the input sequence is monotone decreasing, then our algorithms can compete with a prophet who knows the realization of the sequence in advance. However, in general, it is not possible to compete against such a strong benchmark, since the gap between the two benchmarks can be arbitrarily large. Thus, it is only reasonable to compete with an algorithm that knows the structure of the feedback in advance. We formalize the discussion as the following theorem, whose proof is in Appendix~\ref{sec pf benchmark}.

\begin{theorem}\label{th benchmark}
    No algorithm is competitive against $\E \min_i \left(i+X_i\right)$ for the super-martingale stopping problem.
\end{theorem}

\subsection{On the Robustness of Algorithm~\ref{alg detstop} and Algorithm~\ref{alg randstop}}
In this part, we consider the robustness of Algorithm~\ref{alg detstop} and Algorithm~\ref{alg randstop}. Back to our motivation, buying information for offline stochastic optimization. In the model of buying information for offline stochastic optimization, we assume that given a stochastic optimization problem, the learner can solve the problem exactly. However, since most stochastic optimization problems are NP-hard, usually, the learner might only have an $\alpha$-approximate algorithm to solve it. If $X_i$ is the optimal value of the stochastic optimization problem after receiving the $i$th feedback, then the cost to solve the problem for the learner is instead $\Tilde{X_i}$, where $\Tilde{X_i} \in [X_i,\alpha X_i]$. That is to say, if the learner stops at $X_i$, he will pay $i+\Tilde{X_i}$. We remark that in this case, $\Tilde{X_0},\dots,\Tilde{X_n}$ may not satisfies the super-martingale property anymore, thus we cannot apply the analysis of Algorithm~\ref{alg detstop} and Algorithm~\ref{alg randstop} directly. However, we will show that the two algorithms are robust under such perturbation. In other words, Algorithm~\ref{alg detstop} is $2\alpha$-competitive and Algorithm~\ref{alg randstop} is $\frac{e}{e-1}\alpha$-competitive. Formally, we have the following theorem, whose proof is deferred to Appendix~\ref{sec pf robust}.
\begin{theorem}\label{th robust}
    Let $T$ be a tree representation of an instance of the super-martingale stopping problem. Let $\Tilde{T}$ be any tree constructed by changing the value of every leaf $v\in T$ by some value $\Tilde{v} \in [v,\alpha v]$. If we run Algorithm~\ref{alg detstop} over $\Tilde{T}$, then $\alg(\Tilde{T}) \le 2\alpha \opt(T)$ and
    if we run Algorithm~\ref{alg randstop} over $\Tilde{T}$, then $\alg(\Tilde{T}) \le \frac{e}{e-1}\alpha \opt(T)$.
\end{theorem}

\section{Buying Information for Adaptive Stochastic Optimization and Prophet Inequality}\label{sec reduction}

Unlike offline stochastic optimization with feedback, buying information for adaptive stochastic optimization is much more problem-dependent. For this reason, we consider designing competitive learners to buy information for specific problems. We choose Min Sum Set Cover, an extreme case of the adaptive stochastic optimization problem as the first problem studied under the feedback setting.
\begin{definition}[Min Sum Set Cover]
    Let $\B=[n]$ be a set of boxes, each box $i$ contains an unknown number $b_i \in \{0,1\}$. A learner can know $b_i$ by querying box $i$, i.e. the action space $\X=\B$. A scenario $s \in \{0,1\}^n$ is a binary vector that represents the number contained in each box. If scenario $s$ is realized, then for every box $i \in \B$, $s_i=b_i$. A scenario $s$ is covered if a box $i$ such that $s_i=1$ is queried.
    Let $\S$ be a set of scenarios and $\D$ be a probability distribution over $\S$. Let $f$ be a sequence of feedback. A scenario $s^*$ is drawn from $\D$ initially. In each round $t$, a learner takes an action $a_t \in \X$ to query the box $a_t$ and observes the number contained in that box. Given an instance $(\B,\S,\D)$ of Min Sum Set Cover, the goal of a learner is to construct the sequence of boxes $A$ to query to minimize $\E_{s\sim \D}\ell(A,s)$, where $\ell(A,s)$ is the number of boxes in $A$ to query until the drawn scenario $s$ is covered.  
\end{definition}
The main contribution of this section can be broken down into two parts. In the first part, we give a general strategy to shrink the action space of buying information for a broad class of stochastic optimization problems. For such a class of problems, we show that if an $\alpha$-prophet inequality exists for an adaptive stochastic optimization problem with time dependent feedback, which we will define later, then there is a $2\alpha$-competitive learner for buying information for adaptive stochastic optimization. In the second part, using such an idea, we construct an $8$-competitive learner to buy information for Min Sum Set Cover(MSSC) by showing a 4-prophet inequality for MSSC with time dependent feedback. Furthermore, we will establish information theoretic lower bounds for MSSC under both settings.

\subsection{Time Dependent Feedback and Prophet Inequality}\label{sec time dependent}
A prophet inequality for an adaptive stochastic optimization is established when a signal arrives from $f_t$ for free in each round. Formally, we have the following model.

\begin{definition}[Adaptive Stochastic Optimization with Time Dependent Feedback]
    Let $(\X,\S,\ell,\D)$ be an adaptive stochastic optimization problem. $\F=\{f_t\}_{t=0}^\infty$ be a sequence of feedback.
     Initially, a scenario $s$ is drawn according to $\D$. In each time round $t$, a learner receives a signal $y_t(s)$ from $f_t(s)$, then takes an action $a_t \in \X$.
 An adaptive stochastic optimization problem with time dependent feedback $(\X,\S,\ell,\D,\F)$ is to make decisions to construct a sequence of actions $A$ adaptively to minimize $\E_{s \sim \D} \ell(A,s)$.
\end{definition}
If we denote by $\cost(\A,I)$ be the expected cost of a learner $\A$ at a given instance $I$, then a learner is $\alpha$-competitive if for every instance $I$, $\cost(\A,I)\le \min_\A \cost(\A,I)$. In particular, here we are competing with a learner who knows $\F$ in advance. We say a stochastic optimization $(\X,\S,\ell,\D)$ satisfies an $\alpha$-prophet inequality if there is an $\alpha$-competitive learner for the corresponding stochastic optimization problem with time dependent feedback.
We have the following theorem to establish the relation between the two problems. 
\begin{theorem}\label{th reduction}
If there is an $\alpha$-competitive learner for Min Sum Set Cover with Time Dependent Feedback, then there is a $2\alpha$-competitive learner for Buying Information for Min Sum Set Cover.
\end{theorem}
Although the statement of Theorem~\ref{th reduction} is on MSSC here, the same results actually hold for a broader class of problems, where the loss function can be written as a covering function. Due to space limitations, we leave the general statement and the proof of Theorem~\ref{th reduction} for Appendix~\ref{sec pf reduction}.

\subsection{Min Sum Set Cover with Time Dependent Feedback}\label{sec mssc}
In this part, we establish a 4-prophet inequality for MSSC with time dependent feedback via the following theorem.
\begin{theorem}\label{th time dependent up}
    Algorithm~\ref{alg greedy}, a simple greedy learner is $4$-competitive for Min Sum Set Cover with Time Dependent Feedback.
\end{theorem}

\begin{algorithm}[H]
		\caption{\textsc{Greedy} ($4$-competitive Learner for MSSC with Time Dependent Feedback)}\label{alg greedy}
		\begin{algorithmic}
		
  \FOR{t=0,1,2,\dots}
  
\STATE Receive scenario set $S_t$ that are consistent with the feedback and outcomes received so far.
\STATE Compute $\Pr(i):=\sum_{s \in S_t:s_i=1}\Pr(s\mid S_t)$
\STATE Query any box $i^* \in \arg\max \{\Pr(i) \mid i \in \B\}$.
\IF {$s^*_{i^*}=1$}
\RETURN
\ENDIF

  \ENDFOR
		\end{algorithmic}
	\end{algorithm}

Here we give an overview of our proof, the whole proof is deferred to Appendix~\ref{sec pf greedy}. Our proof is based on a linear programming approach. Assume the feedback is known in advance, then the problem becomes to assign a box for each node of the feedback tree $T(\F)$ to minimize the average number of boxes used to cover the drawn scenario. This problem can be naturally lower bounded by a linear program, and thus every feasible solution to the dual of the linear program gives a lower bound for $\opt$. We will show that a simple greedy algorithm with no knowledge of $T(\F)$ can be used to construct a feasible solution to the dual program such that the cost of the greedy algorithm is at most a quarter times the dual objective of the solution it constructs.

By Theorem 13 in \cite{feige2004approximating}, we know that for every $\epsilon>0$, it is NP-hard to approximate MSSC within a ratio of $4-\epsilon$. MSSC is a very special case of MSSC with Time Dependent Feedback, thus the result given by Theorem~\ref{th time dependent up} is tight if we only consider learners that can be implemented in poly-time. However, in the classic MSSC, if we allow a learner to be implemented in super-polynomial time, then we can simply compute the optimal order of box to query using a brute force method. This gives a natural question. Is the knowledge of $\F$ useful? We show that such knowledge is indeed useful by giving the following information theoretical lower bound for MSSC with Time Dependent Feedback. That is to say, we consider all learners regardless of their running time. We establish the following information theoretic lower bound for MSSC with time dependent feedback. The proof is deferred to Appendix~\ref{sec pf time dependent lb}.

\begin{theorem}\label{th time dependent lb}
For every $\epsilon>0$, there is no deterministic learner that is $2-\epsilon$-competitive for Min Sum Set Cover with Time Dependent Feedback.
\end{theorem}

\subsection{Buying Information for Min Sum Set Cover}\label{sec mssc cost}

In the last section, we establish a prophet inequality for MSSC.
In this section, we go back to the original motivation of buying feedback for adaptive stochastic optimization to discuss the upper bound and information theoretic lower bound for MSSC when asking for feedback requires some cost. The model of the problem is given as follows.

According to Theorem~\ref{th time dependent up} and Theorem~\ref{th reduction}, we can immediately obtain an efficient competitive learner to buy feedback for Min Sum Set Cover, which is described in Algorithm~\ref{alg coin}.

\begin{theorem}\label{th costly up}
There is a poly-time learner that is $8$-competitive for buying information for Min Sum Set Cover.
\end{theorem}

	\begin{algorithm}
		\caption{\textsc{GreedyBuying} ($8$-competitive learner for MSSC with Feedback)}\label{alg coin}
		\begin{algorithmic}
		
  \FOR{$t=0,1,2,\dots$}
\STATE Receive scenario set $S_t$ that are consistent with the feedback and outcomes received so far.
\STATE Receive the cost $c_t$ to receive a signal $y_{t+1}$ from $f_{t+1}$
\FOR{$j=1\dots,c_t$}
\STATE Compute $\Pr(i):=\sum_{s \in S_t:s_i=1}\Pr(s \mid S_t)$.
\STATE Query any box $i^* \in \arg\max\{\Pr(i) \min i \in \B\}$.
\IF{$s^*_{i^*}=1$}
 \RETURN
\ELSE 
\STATE Update $S_t \gets S_t \cap \{s \in \S \mid s_{i^*}=0\}.$
\ENDIF
\ENDFOR
\STATE Pay $c_t$ to obtain signal $y_{t+1}$ from $f_{t+1}$.

  \ENDFOR
		\end{algorithmic}
	\end{algorithm}

The main goal of this section is to obtain an information theoretical lower bound for
buying information for MSSC. We establish the information theoretic lower bound via the following theorem, whose proof is in Appendix~\ref{sec pf costly lb}.
\begin{theorem}\label{th costly lb}
For every $\epsilon>0$, there is no deterministic algorithm that is $2-\epsilon$-competitive for buying information for Min Sum Set Cover.
\end{theorem}

\section{Acknowledgements}
This work was supported by the
NSF Award CCF-2144298 (CAREER).

\nocite{langley00}

\bibliography{example_paper}
\bibliographystyle{icml2023}

\newpage
\appendix
\onecolumn

\section{Equivalence of Randomized and Deterministic Signaling Schemes}\label{sec signal}
In our model, it is sufficient to study the case when each signaling scheme is deterministic. In this part, we give a brief discussion on the equivalence of randomized and deterministic signaling schemes.

Given a set of scenarios $\S$, a distribution $\D$ over $\S$, and a randomized signaling scheme $f$. We show we can construct a modified triple $(\S',\D',f')$ such that $f'$ is a deterministic signaling scheme and $(\S',\D',f')$ is equivalent to $(\S,\D,f)$. The triple is constructed in the following way. $\S'$ contains multiple copies for each $s \in \S$. $\D'$ is a uniform distribution over $\S'$. For every $s \in \S$, assume the range of $f(s)$ is $\{y_1(s),\dots,y_k(s)\}$ and the set of copies is $\{Y_1(s),\dots,Y_k(s)\}$ accordingly. The sizes of the copies are made such that if we draw a scenario according to $\D'$, the probability that it is a copy of $s$ is equal to the probability of obtaining $s$ from $\D$. Furthermore, if we uniformly draw a copy from 
$\{Y_1(s),\dots,Y_k(s)\}$ the probability that we obtain a copy from $Y_i(s)$ is equal to the probability that we receive $y_i(s)$ from $f(s)$. In this way, we define $f'(s')=y_i(s)$ if $s' \in Y_i(s)$. Thus, we obtain an equivalent triple $(\S',\D',f')$ with a deterministic signaling scheme.

\section{Equivalence of Super-Martingale Stopping and Buying Information for Offline Stochastic Optimization} \label{sec formulation}
In this part, we give a brief discussion on the equivalence of the super-martingale stopping problem and buying information for offline stochastic optimization problems.

 We have seen that the super-martingale stopping problem is a special case of buying information for offline stochastic optimization. To see the other direction, it remains to see that given an instance $I=(\X,\S,\ell,\D,\F,\C)$ of buying information for offline stochastic optimization, we can assume each $c_t=1 \in \C$. We give the intuition here via the definition of feedback tree. Let $T(\F)$ be a feedback tree. Assume a learner arrives at a node $v$ of $T(\F)$, the posterior distribution of the stochastic optimization problem at $v$ is $D_v$ and the cost to move to the next node $v'$ is $c_v$. Then we can add $c_v-1$ virtual nodes between $v$ and $v'$ such that the posterior distribution at each node is $D_v$ and the cost to move to the next node is $1$. After the modification, we can run any algorithm for the super-martingale stopping problem over the modified instance. We pay $c_v$ to move to $v'$ if and only if we reach $v'$ in the modified instance. In this way, any $\alpha$-competitive algorithm for the super-martingale stopping problem can be used to construct an $\alpha$-competitive learner to buy information for offline stochastic optimization problems.

\section{Natural Algorithms Fail for Martingale Stopping Problem}\label{sec counter}

In this section, we show some natural algorithms that work for ski-rental problems but fail for the super-martingale stopping problem. According to \cite{karlin1994competitive}, it is well-known that the following algorithm is $2$-competitive for the ski-rental problem.

	\begin{algorithm}[H]
		\caption{\textsc{ClassicSkiRental} ($2$-competitive deterministic algorithm for ski-rental)}\label{alg skistop}
		\begin{algorithmic}
  \FOR{$i=0,1,2,\dots$}
  
\STATE Observe $X_i=v_i$ 
\IF{$v_i \le i$} 
\STATE Stop and pay $i+v_i$.
\ENDIF
  
  \ENDFOR
		\end{algorithmic}
	\end{algorithm}

\begin{theorem}
    Algorithm~\ref{alg skistop} is not competitive for the super-martingale stopping problem.
\end{theorem}

\begin{proof}
We construct a sequence of instance $I_n$ of the super-martingale stopping problem. Denote by $\alg(I_n)$ the cost of Algorithm~\ref{alg skistop} over instance $I_n$ and denote by $\opt(I_n)$ the optimal cost of $I_n$. We will show that $\alg(I_n) \ge H_n \opt(I_n)$, where $H_n$ is the $n$th harmonic number.

Let $(X_0,\dots,X_n)$ be the sequence of random variables for instance $I_n$. Define $X_0=1$ to be a constant. For every $i\ge 1$, $X_i$ can take two possible values. Given $X_{i-1}$, $X_i=0$ with probability $\frac{1}{i+1}$ and with probability $\frac{i}{i+1}$, $X_i=\frac{i+1}{i} X_{i-1}$. That is to say, $X_i$ is either $0$ or $i+1$ and $\E X_i=1$.

Assume we run Algorithm~\ref{alg skistop} over instance $I_n$. Suppose we just observe $X_i$. If $X_i=0$, then we stop and pay $i$ right away. If $X_i=i+1$, then Algorithm~\ref{alg skistop} will keep querying $X_{i+1}$. Denote by $i^*$ the random variable of the stopping time of Algorithm~\ref{alg skistop}. Then, we have 
\begin{align*}
    \E i^* = \sum_{i=1}^n \frac{i}{i+1}\prod_{j=1}^{i-1}\frac{j}{j+1}= \sum_{i=1}^n \frac{1}{i+1} = H_n-1. 
\end{align*}
On the other hand, we know from the construction of the instance that $\E X_{i^*}=1$, since $\E X_i=1$ for every $i$. Thus the total cost of the algorithm is $\alg(I_n) = \E i^*+X_{i^*} = H_n$. On the other hand, we have $\opt(I_n) \le 1$, since it can simply stop at the beginning. This gives $\alg(I_n) \ge H_n \opt(I_n)$, which implies that Algorithm~\ref{alg skistop} is not competitive.

\end{proof}

The reason why Algorithm~\ref{alg skistop} fails is that $(X_1,\dots,X_n)$ might be an increasing sequence, which forces the algorithm to keep querying the next box forever. To avoid keeping querying boxes forever, a natural idea is to change the stopping rule by looking at the smallest value we have seen so far. However, it turns out that such a stopping rule still fails. We consider the following algorithm.

	\begin{algorithm}[H]
		\caption{\textsc{RevisedSkiRental} ($2$-competitive deterministic algorithm for ski-rental)}\label{alg minstop}
		\begin{algorithmic}
  \FOR{$i=0,1,2,\dots$}
  
\STATE Observe $X_i=v_i$ and set $v^*=\min \{v_0,\dots,v_i\}$. 
\IF{$v^* \le i$} 
\STATE Stop and pay $i+v_i$.
\ENDIF
  
  \ENDFOR
		\end{algorithmic}
	\end{algorithm}

\begin{theorem}
    Algorithm~\ref{alg minstop} is not competitive for the super-martingale stopping problem.
\end{theorem}

\begin{proof}
     We construct a sequence of instance $I_n$ of the super-martingale stopping problem. Denote by $\alg(I_n)$ the cost of Algorithm~\ref{alg minstop} over instance $I_n$ and denote by $\opt(I_n)$ the optimal cost of $I_n$. We will show that $\alg(I_n) \ge \Omega(n) \opt(I_n)$.

     Let $(X_0,\dots,X_n,X_{n+1})$ be the sequence of random variables of instance $I_n$ of the super-martingale stopping problem. Define $X_0=n$ and $X_{n+1}=0$. For $i \in [n]$, $X_i$ can take two possible values. Given $X_{i-1}$, $X_i=e^nX_{i-1}$ with probability $e^{-n}$ and $X_i=0$ with probability $1-e^{-n}$. That is to say for $i \in [n]$, $\E X_n = n$. Notice that according to the stopping rule of Algorithm~\ref{alg minstop}, $X_{n+1}$ will never be queried by the algorithm. Thus, we have $\alg(I_n) \ge \E X_i =n.$

     On the other hand, we consider an algorithm that keeps querying $X_{i+1}$ if $X_i \neq 0$. Denote by $i^*$ the stopping time of this algorithm. We know that $\E X_{i^*} =0$. Furthermore, we have 
     \begin{align*}
         \E i^* = \sum_{i=1}^{n+1}i(1-e^{-n})\prod_{j=1}^{i-1}e^{-n} \le e^{-n}\sum_{i=1}^{n+1}i \in O(1).
     \end{align*}
This implies that $\opt(I_n) \le \E i^*+X_{i^*} \in O(1)$, while $\alg(I_n) \in \Omega(n)$. Thus, Algorithm~\ref{alg minstop} is not competitive.
     
\end{proof}

\section{A Simple Randomized Algorithm for Martingale Stopping Problem}\label{sec simple}

In this section, we give a simple randomized $2$-competitive algorithm for the super-martingale stopping problem.

	\begin{algorithm}[H]
		\caption{\textsc{ThrowCoin} (Simple $2$-competitive randomized algorithm for super-martingale stopping)}\label{alg throw coin}
		\begin{algorithmic}
  \FOR{$i=0,1,2,\dots$}
\STATE Observe $X_i=v_i$.
\STATE Stop and pay $i+v_i$ with probability $\min\{1, 1/v_i\}$.
  \ENDFOR
		\end{algorithmic}
	\end{algorithm}

\begin{theorem}
    Algorithm~\ref{alg throw coin} is 2-competitive for the super-martingale stopping problem.
\end{theorem}

\begin{proof}
    Let $(X_0,\dots,X_n)$ be a sequence of random variables, and let $T$ be the tree representation of the sequence. Denote by $\opt(T)$ the optimal cost of the instance and denote by $\alg(T)$ the cost of Algorithm~\ref{alg throw coin} over the instance. We prove the theorem using inductions on the number of random variables, which is also the depth of $T$.
    
    If $\dep(T)=0$, which means there is only one random variable $X_0$ in the sequence, the cost of any algorithm is $X_0$ and the theorem holds trivially. Assuming the theorem holds for any tree with depth $n-k$, we show the theorem holds for any tree with depth $n-k-1$. Let $T$ be a tree of an instance of super-martingale stopping problem such that $\dep(T)=n-k-1$. Let $v$ be the root of $T$ and let ${v^1,\dots,v^k}$ be the children of $v$. Denote by $T^i$ the subtree rooted at $v^i$. By a dynamic programming approach, we know that 
    \begin{align*}
        \opt(T) = \min \{v,1+\sum_{i=1}^k \Pr(v^i)\opt(T^i)\}.
    \end{align*}
We consider two cases. In the first case, $\opt(T)=v$. Without loss of generality, we assume $v>1$, otherwise, the algorithm will simply stop at $v$. The cost of Algorithm~\ref{alg throw coin} is 
\begin{align*}
    \alg(T) & = \frac{1}{v}v+(1-\frac{1}{v})(1+\sum_{i=1}^k\Pr(v^i)\alg(T^i)) \\
    & \le \frac{1}{v}v+(1-\frac{1}{v})(1+2\sum_{i=1}^k\Pr(v^i)\opt(T^i)) \\
    & \le \frac{1}{v}v+(1-\frac{1}{v})(1+2\sum_{i=1}^k\Pr(v^i)v^i) \\
    & \le 1+1-\frac{1}{v}+2v-2 \le 2v.
\end{align*}
Here, in the first inequality, we use the induction hypothesis, in the third inequality, we use the super-martingale property.

In the second case, $\opt(T) = 1+\sum_{i=1}^k \Pr(v^i)\opt(T^i)$. Similarly, we have 
\begin{align*}
    \alg(T) & = \frac{1}{v}v+(1-\frac{1}{v})(1+\sum_{i=1}^k\Pr(v^i)\alg(T^i)) \\
    & \le \frac{1}{v}v+(1-\frac{1}{v})(1+2\sum_{i=1}^k\Pr(v^i)\opt(T^i)) \\
    & \le 2\left( 1+\sum_{i=1}^k \Pr(v^i)\opt(T^i) \right).
\end{align*}
This shows that for every instance with a tree representation $T$, $\alg(T) \le 2\opt(T)$. This implies Algorithm~\ref{alg throw coin} is 2-competitive.

\end{proof}

\section{Miss Proof in Section~\ref{sec martingale}}

\subsection{Proof of Lemma~\ref{lm up time}}\label{sec pflm up time}

\begin{proof}
    We prove this lemma using induction on the depth of $v$. If $v$ has a depth of $n$ ($v$ is a leaf), then Lemma~\ref{lm up time} follows directly by Lemma~\ref{lm lb time}, since $Q_{p_j}^{-1}(\rho)-Q_{p_j}^{-1}(\rho^*) = \int_{\rho^*}^{\rho}vdw=(\rho-\rho^*)v.$ Assume Lemma~\ref{lm up time} holds for every node $v'$ with depth $n-k$, we show this for a node $v$ with depth $n-k-1$. We notice that if $\rho \le Q_{p_j}(n-k)$, then this is correct by Lemma~\ref{lm lb time}. So in the rest of the proof, we assume $\rho > Q_{p_j}(n-k)$.
    Let $\{u_j\}_{j=1}^\ell$ be the set of children of $v$ and let $D(u_j) \subseteq \{p_j\}_{j=1}^k$ be the set of paths that passes $u_j$. 
    then we have 
    \begin{align*}
        \sum_{j=1}^k\Pr(p_j)(Q_{p_j}^{-1}(\rho)-Q_{p_j}^{-1}(\rho^*)) 
        & = \sum_{j=1}^k\Pr(p_j)(n-k-Q_{p_j}^{-1}(\rho^*))+\sum_{j=1}^\ell\sum_{p \in D(u_j)}\Pr(p)(Q_{p_j}^{-1}(\rho)-(n-k)) \\
        & \le \sum_{j=1}^k\Pr(p_j)(Q_{p_j}(n-k)-\rho^*)v+\sum_{j=1}^\ell\Pr(u_j)(\rho-(Q_p(n-k)))u_j \\
        & = \Pr(v)(Q_{p}(n-k)-\rho^*)v+\sum_{j=1}^\ell\Pr(u_j)(\rho-(Q_p(n-k)))u_j \\
        & \le \Pr(v)(Q_{p}(n-k)-\rho^*)v+\Pr(v)(\rho-(Q_p(n-k)))v \\
        & = \Pr(v)(\rho-\rho^*)v.
    \end{align*}
Here, in the first inequality, we use the assumption of induction and in the second inequality, we use the fact that $\sum_{j=1}^\ell \Pr(u_j \mid v)u_j \le u_j$.
\end{proof}

\subsection{Proof of Theorem~\ref{th martingale det}}\label{sec pf martingale det}

\begin{proof}
We show Algorithm~\ref{alg detstop} is $2$-competitive.
    Let $T$ be a tree representation of an instance of the super-martingale stopping problem. We maintain two sets of nodes $\Tilde{O}$ and $\Tilde{A}$ in the following way. For each path $p \subseteq T$. We travel down $p$ from the root of $T$ and stop traveling at a node $v$ of $p$ if either $v$ is a stopping node of $\opt(T)$ or it is a stopping node of Algorithm~\ref{alg detstop}. In the first case, we add $v$ to $\Tilde{O}$, otherwise, we add it to $\Tilde{A}$. We denote by $\Tilde{T}$ the subtree of $T$ with the set of leaves $\Tilde{A} \cup \Tilde{O}$. Furthermore, let $P(v)$ be the path of $\Tilde{T}$ that ends at $v \in \Tilde{O}\cup \Tilde{A}$.
    Then we have the following lower bound for $\opt(T)$.
    \begin{align*}
        \opt(T) \ge \sum_{v \in \Tilde{O}}\Pr(v)\left(v+\dep(v) \right)+\sum_{v\in\Tilde{A}}\Pr(v)\dep(v)
    \end{align*}

To upper bound $\alg(T)$, we will need to establish the following inequality and claim.   
Let $P$ be a path such that there is some $v \in P \cap \Tilde{O}$, then there must be some stopping node $f_P(v) \in P$ of $\alg(T)$ that has $v$ as its ancestor. Let $D(v)$ be the set of paths that passes $v$. We know from the stopping rule of Algorithm~\ref{alg detstop} that for every $P \in D(v)$, $Q_P(\dep(f_P(v))) \le 1$.
By Lemma~\ref{lm up time}, we have 
\begin{align}\label{eq time up}
    \sum_{P \in D(v)}\Pr(P)(\dep(f_P(v))-\dep(v))\le \sum_{P \in D(v)}\Pr(P)(Q_P^{-1}(1)-\dep(v))  \le \Pr(v)\left(1-Q_P(v)\right)v \le \Pr(v)v.
\end{align}

Furthermore, we next prove the following claim.
\begin{claim}\label{cl feedback}
Let $T'$ be a subtree of $T$ with the same root of $T$. Let $S$ be the set of leaves of $T'$. For each $v\in S$, denote by $P(v)$ the path from the root to $v$. If every path of $T$ has a node in $S$, then
    $ \sum_{v \in S}\Pr(v)vQ_{P(v)}(\dep(v)) \le  \sum_{v \in S}\Pr(v)\dep(v) $.
\end{claim}

\begin{cpf}
    Let $v \in S$ be a leave of $T'$. Assume that $\dep(v)=i$ and $P(v)=(v_0,\dots,v_i)$. Then 
    \begin{align}\label{eq Q}
        vQ_{P(v)}(i)=v_i\sum_{j=0}^{i-1}\frac{1}{v_j}.
    \end{align}
    Now we prove this claim by induction on the depth of $T'$. If $T'$ has a depth of  $0$, then the claim holds trivially. Now we assume the claim for any tree with depth $k$, we show this holds for a tree $T'$ with depth $k+1$. We remove the nodes with depth $k+1$ in $T'$ and denote by the remaining tree $\Bar{T}$. Denote by $\Bar{S}$ the leaves of $\Bar{T}$ and denote by $K$ the set of leaves of $\Bar{T}$ with depth $k$. For every node $v$, let $N(v)$ be the set of children of $v$. Then we have 
    \begin{align*}
        \sum_{v \in S}p(v)vQ_{P(v)}(\dep(v)) & = \sum_{v \in \Bar{S}}\Pr(v) vQ_{P(v)}(\dep(v)) \\
        &+ \sum_{v \in K}\Pr(v)\sum_{u \in N(v)}\left(\Pr(u \mid v)(uQ_{P(u)}(k+1)-vQ_{P(v)}(k))\right) \\
        & \le \sum_{v \in \Bar{S}}\Pr(v)\dep(v) + \sum_{v \in K}\Pr(v)\sum_{u \in N(v)}\left(\Pr(u \mid v)(uQ_{P(u)}(k+1)-vQ_{P(v)}(k))\right) \\
        & \le \sum_{v \in \Bar{S}}\Pr(v)\dep(v) + \sum_{v \in K}\Pr(v)\sum_{u \in N(v)}\left(\Pr(u \mid v)u(Q_{P(u)}(k+1) - Q_{P(v)}(k))\right) \\
        & = \sum_{v \in \Bar{S}}\Pr(v)\dep(v) + \sum_{v \in K}\Pr(v) = \sum_{v \in S}\Pr(v)\dep(v).
    \end{align*}
Here, the first inequality follows by our induction, the second inequality follows by the super-martingale property and the second equality follows by \eqref{eq Q}.
    
\end{cpf}

This gives the following upper bound for $\alg(T)$.
\begin{align*}
    \alg(T) &\le \sum_{v \in \Tilde{O}} \left(\Pr(v)\left(v+\dep(v)\right)+ \sum_{P \in D(v)}\Pr(P)(\dep(f_P(v))-\dep(v))\right) + \sum_{v \in \Tilde{A}}\Pr(v)\left(v+\dep(v)\right)\\
    & \le \sum_{v \in \Tilde{O}}\Pr(v) \left(\dep(v)+2v\right) + \sum_{v \in \Tilde{A}}\Pr(v)\left(v+\dep(v)\right) \\
    & \le \sum_{v \in \Tilde{O}}\Pr(v) \left(\dep(v)+2v\right) + \sum_{v \in \Tilde{A}}\Pr(v)\left(v(Q_{P(v)}(\dep(v))+1)+\dep(v)\right) \\
    & \le \sum_{v \in \Tilde{O}}\Pr(v) \left(\dep(v)+2v\right) + \sum_{v \in \Tilde{A}}\Pr(v)\left(v(Q_{P(v)}(\dep(v))+1)+ \dep(v) \right) \\
    &+ \sum_{v \in \Tilde{O}}\Pr(v)\left(vQ_{P(v)}(\dep(v)) \right) \\
    & \le 2\sum_{v \in \Tilde{O}}\Pr(v)v+2\sum_{v \in \Tilde{A}}\Pr(v)v + 2 \sum_{v\in\Tilde{A}}\Pr(v)\dep(v)+2\sum_{v\in\Tilde{A}}\Pr(v)\dep(v) \\
    & \le 2\opt(T).
\end{align*}
Here, in the first inequality, we used the super-martingale property of $T$. In the second inequality, we use \eqref{eq time up}. In the third inequality, we use the stopping rule of Algorithm~\ref{alg detstop}. The second last inequality follows by Claim~\ref{cl feedback}.

\end{proof}

\subsection{Proof of Theorem~\ref{th martingale rand}}\label{sec pf martingale rand}

\begin{proof}
We show Algorithm~\ref{alg randstop} is $e/(e-1)$-competitive.    
Let $T$ be a representation of an instance of the super-martingale stopping problem. Let $S=\{v^*_j\}_{j=1}^k$ be the set of stopping nodes of $\opt(T)$. Let $u \in S$ and let $P(u) \subseteq T$ be the path from the root to $u$. 
We notice that we can assume the depth of $u$ is at most $Q^{-1}_{P(u)}(1)$. Since the cost of Algorithm~\ref{alg randstop} only depends on the value of nodes with depth strictly less than $Q^{-1}_{P(u)}(1)$, we can assume every node with a depth larger than $Q^{-1}_{P(u)}(1)$ has a value of $0$. This assumption doesn't affect the cost of Algorithm~\ref{alg randstop} but will force every $u\in S$ has depth at most $\lceil Q^{-1}_{P(u)}(1)\rceil$. Under this assumption, if a node $u$ has depth exactly $\lceil Q^{-1}_{P(u)}(1)\rceil$, we can furthermore assume the contribution of $u$ to the cost of $\opt(T)$ is $Q^{-1}_{P(u)}(1)$. This will only decrease the cost of $\opt(T)$. So in the rest of the proof, every $u$ in $S$ has a depth at most $Q^{-1}_{P(u)}(1)$. In particular, this implies for every $u \in S$, there exists some $\rho_{u} \in [0,1]$ such that $Q_{P(u)}(\dep(u)) = \rho_{u}$. Thus, we can write 
\begin{align*}
    \opt(T) = \sum_{u \in S}\Pr(u) \left( u + Q^{-1}_{P(u)}(\rho_u)  \right).
\end{align*}
On the other hand, we can decompose the cost of the algorithm according to $S$. For every $u \in S$, we define $D(u)$ to be the set of paths in $T$ from the root to a leaf that passes $u$. Then we can write the cost of the algorithm
\begin{align*}
    \alg(T) & \le \sum_{u \in S}\Pr(u)\sum_{p'\in D(u)}\Pr(p'\mid u)\int_0^1\left(v_{p'}(Q_{p'}^{-1}(\rho))+Q_{p'}^{-1}(\rho)\right)p(\rho)d\rho,
\end{align*}
where we use the fact that when the algorithm stops at time $t$, the depth of the stopping node is at most $t$. This implies 
\begin{align*}
    \alg(T)-\opt(T) & \le \sum_{u \in S}\Pr(u) \int_0^1\sum_{p'\in D(u)}\Pr(p'\mid u)\left[\left(v_{p'}(Q_{p'}^{-1}(\rho))+Q_{p'}^{-1}(\rho)\right)-\left( u + Q^{-1}_{P(u)}(\rho_u)  \right)\right]p(\rho)d\rho \\
    & = \sum_{u \in S}\Pr(u) \int_0^{\rho_u}\sum_{p'\in D(u)}\Pr(p'\mid u)\left[\left(v_{p'}(Q_{p'}^{-1}(\rho))+Q_{p'}^{-1}(\rho)\right)-\left( u + Q^{-1}_{P(u)}(\rho_u)  \right)\right]p(\rho)d\rho \\
    &- \sum_{u \in S}\Pr(u) \int_{\rho_u}^{1}\sum_{p'\in D(u)}\Pr(p'\mid u)\left[\left( u + Q^{-1}_{P(u)}(\rho_u)  \right)-\left(v_{p'}(Q_{p'}^{-1}(\rho))+Q_{p'}^{-1}(\rho)\right)\right]p(\rho)d\rho \\
    & \le \sum_{u \in S}\Pr(u) \int_0^{\rho_u}\sum_{p'\in D(u)}\Pr(p'\mid u)\left[\left(v_{p'}(Q_{p'}^{-1}(\rho))+Q_{p'}^{-1}(\rho)\right)-\left( u + Q^{-1}_{P(u)}(\rho_u)  \right)\right]p(\rho)d\rho \\
    & +\sum_{u \in S}\Pr(u) \int_{\rho_u}^{1}\sum_{p'\in D(u)}\Pr(p'\mid u)\left( Q_{p'}^{-1}(\rho) - Q^{-1}_{P(u)}(\rho_u)\right)p(\rho)d\rho \\
    & \le \sum_{u \in S}\Pr(u) \int_0^{\rho_u}\sum_{p'\in D(u)}\Pr(p'\mid u)\left[\left(v_{p'}(Q_{p'}^{-1}(\rho))+Q_{p'}^{-1}(\rho)\right)-\left( u + Q^{-1}_{P(u)}(\rho_u)  \right)\right]p(\rho)d\rho \\
    & +\sum_{u \in S}\Pr(u) \int_{\rho_u}^{1}(\rho-\rho_u)up(\rho)d\rho \\
    & = \sum_{u \in S}\Pr(u) \int_0^{\rho_u}\left(v_{P(u)}(Q_{P(u)}^{-1}(\rho))+Q_{P(u)}^{-1}(\rho)- Q^{-1}_{P(u)}(\rho_u)  \right)p(\rho)d\rho \\
    & + \sum_{u \in S}\Pr(u) \left( \int_{\rho_u}^{1}(\rho-\rho_u)up(\rho)d\rho -\int_0^{\rho_u}up(\rho)d\rho\right).
\end{align*}
Here, the second inequality follows the super-martingale property of the sequence of random variables starting from node $u$. The second inequality follows by Lemma~\ref{lm up time}.

Recall our goal is to show that $\alg(T)-\frac{e}{e-1}\opt(T) = \alg(T)-\opt(T)-\frac{1}{e-1}\opt(T) \le 0$. For every $u \in S$, we define two functions $F_u(s)$ and $G_u(\tau)$ as follows. Let
\begin{align*}
    F_u(s) := \int_0^{s}\left(v_{P(u)}(Q_{P(u)}^{-1}(\rho))+Q_{P(u)}^{-1}(\rho)- Q^{-1}_{P(u)}(s)  \right)p(\rho)d\rho - \frac{1}{e-1}Q^{-1}_{P(u)}(s)
\end{align*}
and
\begin{align*}
    G_u(\tau) := \int_{\rho_u}^{1}(\rho-\rho_u)\tau p(\rho)d\rho -\int_0^{\rho_u}\tau p(\rho)d\rho - \frac{1}{e-1}\tau.
\end{align*}
From our above discussion, we know that 
\begin{align*}
    \alg(T)-\frac{e}{e-1}\opt(T) = \alg(T)-\opt(T)-\frac{1}{e-1}\opt(T) \le \sum_{u \in S}\Pr(u)\left(F_u(\rho_u)+G_u(u)\right).
\end{align*}
It is sufficient to show for every $u$, $F_u(s) \le 0, \forall s \in [0,\rho_u]$ and $G_u(\tau) \le 0, \forall \tau \ge 0$. We first look at $F_u(s)$. Recall the definition of the density function is $p(\rho)= \frac{e^\rho}{e-1}$. We know from Lemma~\ref{lm lb time} that 
\begin{align*}
F_u(s) & = \int_0^s \left(v_{P(u)}(Q_{P(u)}^{-1}(\rho)) - \int_\rho^sv_{P(u)}(Q_{P(u)}^{-1}(w)dw   \right)p(\rho)d\rho - \frac{1}{e-1}\int_0^sv_{P(u)}(Q_{P(u)}^{-1}(w)dw \\
& = \int_0^rv_{P(u)}(Q_{P(u)}^{-1}(w)dw\frac{e^r}{e-1}\bigg|_{r=0}^{r=s}-\int_0^s\int_0^\rho v_{P(u)}(Q_{P(u)}^{-1}(w)dwp(\rho)d\rho\\
&-\int_0^s\int_\rho^sv_{P(u)}(Q_{P(u)}^{-1}(w)dwp(\rho)d\rho- \frac{1}{e-1}\int_0^sv_{P(u)}(Q_{P(u)}^{-1}(w)dw \\
& =\frac{e^s}{e-1}\int_0^sv_{P(u)}(Q_{P(u)}^{-1}(w)dw-\int_0^s\int_0^sv_{P(u)}(Q_{P(u)}^{-1}(w)dwp(\rho)d\rho - \frac{1}{e-1}\int_0^sv_{P(u)}(Q_{P(u)}^{-1}(w)dw \\
& = \left( \frac{e^s}{e-1}-\frac{e^s-1}{e-1}-\frac{1}{e-1}\right)\int_0^sv_{P(u)}(Q_{P(u)}^{-1}(w)dw =0.
\end{align*}
Then we look at $G_u(\tau)$. We have 
\begin{align*}
    \frac{dG_u(\tau)}{d\tau} & = \int_{\rho_u}^1(\rho-\rho_u)p(\rho)d\rho-\int_0^{\rho_u}p(\rho)d\rho-\frac{1}{e-1} \\
    & = \frac{\rho e^\rho}{e-1}\bigg|_{\rho=\rho_u}^{\rho=1}-\int_{\rho_u}^1p(\rho)d\rho -\frac{\rho_u(e-e^{\rho_u})}{e-1}-\int_0^{\rho_u}p(\rho)d\rho-\frac{1}{e-1} \\
    & = -\frac{\rho_ue}{e-1}\le 0.
\end{align*}
This implies that $G_u(\tau) \le G_u(0)=0$. Put the above arguments together, we obtain $\alg(T) \le \frac{e}{e-1}\opt(T)$.
\end{proof}

\subsection{Proof of Theorem~\ref{th benchmark}}\label{sec pf benchmark}

\begin{proof}
    Let $I$ be an instance of super-martingale stopping problem. Let $\opt(I)=\min \E\left(i^*+X_{i^*}\right)$ and $\opt'= \E \min_i \left(i+X_i\right)$. We will construct a sequence of instance $I_N$ such that $\opt(I_N) \ge \Omega(N) \opt'(I_N)$, showing that the gap between the two benchmarks can be arbitrarily large.

    Let $(X_0,X_1,\dots)$ be the sequence of random variables of instance $I_N$. Define $X_0=N$. For every $i \ge 1$, $X_i$ can take two possible values. Given $X_i$, $X_{i+1}=e^NX_i$ with probability $e^{-N}$ and $X_{i+1}=0$ with probability $1-e^{-N}$. That is to say, the sequence of random variables is a super-martingale with a mean equal to $N$. Thus, the optimal stopping rule is to simply stop at $X_0$ and $\opt(I_N)=N$. On the other hand, consider any realization $(x_0,x_1,\dots)$ of the sequence. We notice from the construction that if $x_i=0$ then for every $j>i$, $x_j=0$. Denote by $i'$ the smallest index such that $x_{i'}=0$. Then we have $\min i+x_i = i'$ if $i' \le N$ and $\min i+x_i = N$ if $i'>N$. Since $\Pr(i'=i)=(1-e^{-N})e^{-(i-1)N}$, we have
    \begin{align*}
        \opt'(I_N) = \sum_{i=1}^N i(1-e^{-N})e^{-(i-1)N} + \sum_{i=N+1}^\infty N(1-e^{-N})e^{-(i-1)N} \in O(1).
    \end{align*}
This implies $\opt(I_N) \ge \Omega(N) \opt'(I_N)$.
    
\end{proof}

\subsection{Proof of Theorem~\ref{th robust}}\label{sec pf robust}

\begin{proof}

It is sufficient to show that if we run Algorithm~\ref{alg detstop} or Algorithm~\ref{alg randstop} then $\alg(\Tilde{T}) \le \alpha \alg(T)$. Recall that the only difference between Algorithm~\ref{alg detstop} and Algorithm~\ref{alg randstop} is that they use different threshold $\rho$. Algorithm~\ref{alg detstop} uses $\rho=1$ and Algorithm~\ref{alg randstop} uses a random threshold.
Let $\rho \in [0,1]$ be a realization of the random threshold used in Algorithm~\ref{alg detstop} and Algorithm~\ref{alg randstop}. We denote by $\alg_\rho(\Tilde{T})$ and $\alg_\rho(T)$ the cost of the Algorithm on the corresponding instances with a threshold $\rho$. In the rest of the proof, we will show $\alg_\rho(\Tilde{T}) \le \alpha \alg_\rho(T)$ for every $\rho \in [0,1]$. This will directly imply that $\alg(\Tilde{T}) \le \alpha \alg(T)$.

Since the only difference between $T$ and $\Tilde{T}$ is the value of each node, let $p=(v_0,v_1,\dots,v_n)$ be a path in $T$, we define $\Tilde{v}_p(t) = \Tilde{v_i}$ if $t \in [i,i+1)$. We can also define $\Tilde{Q}_p(t)$ and $\Tilde{Q}_p^{-1}(t)$ in the similar way. Using these notations, we have 
\begin{align*}
    \alg_\rho(T) & = \E_p \left(Q^{-1}_p(\rho) + v_p(Q^{-1}_p(\rho))\right)
                  =  \int_0^\rho \E_p v_p(Q^{-1}_p(w))dw + \E_p v_p(Q^{-1}_p(\rho) 
                  \ge \rho\E_p v_p(Q^{-1}_p(\rho))+\E_p v_p(Q^{-1}_p(\rho)).
\end{align*}
Here, the second equality follows by Lemma~\ref{lm lb time} and the inequality follows by the super-martingale property of $T$.

On the other hand, for every path $p$, since for every $v\in T$, $\Tilde{v} \ge v$, we know that $\Tilde{Q}_p^{-1}(\rho) \ge  Q^{-1}_p(\rho)$. If we denote by $t'=Q^{-1}_p(\rho)$, then this implies there exists some $\rho' \le \rho$ such that $\Tilde{Q}_p(t')=\rho'$. In particular, since $\Tilde{v}_p(t) \le \alpha v_p(t)$ for every $t$, it follows that $\rho' \ge \frac{1}{\alpha} \rho$.
Thus, we can write
\begin{align*}
    \alg_\rho(\Tilde{T}) & = \E_p \left( Q^{-1}_p(\rho)+\Tilde{Q}^{-1}_p(\rho)-\Tilde{Q}^{-1}_p(\rho') +\Tilde{v}_p(\Tilde{Q}^{-1}_p(\rho))  \right) \\
    & = \E_p  Q^{-1}_p(\rho)+ \E_p \int_{\rho'}^\rho  \Tilde{v}_p(\Tilde{Q}^{-1}_p(w))dw + \E_p\Tilde{v}_p(\Tilde{Q}^{-1}_p(\rho))   \\
    & \le \E_p  Q^{-1}_p(\rho) + \alpha (\rho-\rho') \E_p v_p(Q^{-1}_p(\rho))+\alpha \E_p v_p(Q^{-1}_p(\rho)) \\
    & \le \E_p  Q^{-1}_p(\rho) + (\alpha -1)\rho \E_p v_p(Q^{-1}_p(\rho))+\alpha \E_p v_p(Q^{-1}_p(\rho)).
\end{align*}
Here, the equality follows Lemma~\ref{lm lb time}, the first inequality follows by the super-martingale property of the $T$ and the second inequality follows by the fact that $\rho' \ge \frac{1}{\alpha} \rho$.
Thus, we obtain
\begin{align*}
    \frac{\alg_\rho(\Tilde{T})}{\alg_\rho(T)} \le \frac{\E_p  Q^{-1}_p(\rho) + (\alpha -1)\rho \E_p v_p(Q^{-1}_p(\rho))+\alpha \E_p v_p(Q^{-1}_p(\rho))}{\E_p Q^{-1}_p(\rho) + \
    \E_p v_p(Q^{-1}_p(\rho))} \le \max \{\alpha,1+(\alpha-1)\frac{\rho \E_p  v_p(Q^{-1}_p(\rho))}{\E_p Q^{-1}_p(\rho)}\}.
\end{align*}
Here we use the fact that if $a,b,c,d \ge 0$, then $\frac{a+b}{c+d} \le \max \{\frac{a}{c},\frac{b}{d}\}$.  By Lemma~\ref{lm lb time} and the super-martingale property, we know that 
\begin{align*}
    \E_p Q^{-1}_p(\rho) = \int_0^\rho \E_p v_p(Q^{-1}_p(w))dw \ge \rho\E_p v_p(Q^{-1}_p(\rho)).
\end{align*}
This implies 
\begin{align*}
    1+(\alpha-1)\frac{\rho \E_p  v_p(Q^{-1}_p(\rho))}{\E_p Q^{-1}_p(\rho)} \le 1+ \alpha-1 = \alpha.
\end{align*}
Thus, $\alg_\rho(\Tilde{T}) \le \alpha \alg_\rho(T)$ for every $\rho \in [0,1]$.

\end{proof}

\section{Missing Proof in Section~\ref{sec reduction}}

\subsection{Proof of Theorem~\ref{th reduction}}\label{sec pf reduction}
As we mentioned in the main body of the paper, Theorem~\ref{th reduction} not only holds for MSSC but also holds for a broader class of stochastic optimization problems.
In this part, we give the general statement and the proof for Theorem~\ref{th reduction}. To begin with, we define a broad class of adaptive stochastic optimization problems for which Theorem~\ref{th reduction} holds.

\begin{definition}[Adaptive Stochastic Optimization with Covering Loss]
    Let $(\X,\S,\ell,\D)$ be an adaptive stochastic optimization problem. For every $s \in \S$, we define a family of sets of actions $C(s) \subseteq 2^\X$. We say a scenario $s$ is covered if a set of actions $A \in C(s)$ are taken. We say the loss function $\ell$ is a covering loss if for every scenario $s \in \S$ and every sequence of actions $\vec a = (a_1,a_2,\dots)$, $\ell(\vec a,s) = \min \{t \mid \exists A \in C(s), \stt A \subseteq\{a_1,\dots,a_t\} \}$, which is the time for $\vec a$ to cover $s$.
\end{definition}
Many adaptive stochastic optimization problems such as MSSC and optimal decision tree problems have covering objective functions. Next, we give a general statement of Theorem~\ref{th reduction}, which builds a connection between adaptive stochastic optimization with time dependent feedback and buying information for adaptive stochastic optimization.

\begin{theorem}[General Version of Theorem~\ref{th reduction}]\label{th reduction restate}
    Let $(\X,\S,\ell,\D)$ be an adaptive stochastic optimization problem with a covering loss function.  If $(\X,\S,\ell,\D)$ satisfies an $\alpha$-prophet inequality, then there is a $2\alpha$-competitive learner for the adaptive stochastic optimization problem with feedback $(\X,\S,\ell,\D,\F,\C)$.
\end{theorem}

\begin{proof}
    Denote by $I=(\X,\S,\ell,\D,\F,\C)$ the instance of buying information for stochastic optimization and $\opt(I)$ be the optimal value of the instance.  Since $(\X,\S,\ell,\D)$ satisfies an $\alpha$-prophet inequality, let $\A$ be an $\alpha$-competitive learner for the stochastic optimization problem with feedback. At time round $t$, denote by $R_t$ the set of outcomes received after taking a set of actions and denote by $Y_t$ a set of signals received from the signaling schemes. Notice that $R_t,Y_t$ are random sets that depend on the random scenarios $s$. Then $a=\A(R_t,Y_t)$ is the next action taken by the learner $\A$. Based on the notations, we design the following algorithm, which will be shown as $2\alpha$-competitive for $I=(\X,\S,\ell,\D,\F,\C)$.

	\begin{algorithm}
		\caption{\textsc{BuyingInformation} ($2\alpha$-competitive algorithm for $(\X,\S,\ell,\D,\F,\C)$)}\label{alg reduction}
		\begin{algorithmic}
		
  \FOR{$t=0,1,2,\dots$}
\STATE Receive $R_{t-1}$ the set of outcomes received so far and $Y_{t-1}$ the signals received so far.
\STATE Receive a cost $c_t$ to receive a signal from $f_{t+1}$.
\FOR{$i=1,\dots,c_t$}
\STATE Take action $a_i=\A(R_{t-1},Y_{t-1})$ and receive outcome $r_i$.

\STATE Update $R_{t-1} \gets R_{t-1} \cup \{r_i\}$.

\IF{$s$ is covered}  \RETURN
\ENDIF

\ENDFOR
\STATE Pay $c_t$ to get signal $y_{t+1}$ from $f_{t+1}$.
\STATE Update $R_t \gets R_{t-1}$, $Y_t \gets Y_{t-1} \cup \{y_{t+1}\}$

\ENDFOR
		\end{algorithmic}
	\end{algorithm}
We notice that during the execution of Algorithm~\ref{alg reduction}, we count the time round in a different way for convenience. This doesn't affect the final cost of the algorithm. We now decompose the cost of Algorithm~\ref{alg reduction} into two parts. Denote by $\A'$ Algorithm~\ref{alg reduction}. For each scenario $s \in \S$, let $\cost(s)$ be the total cost of $\A'$ when $s$ is drawn. We write 
\begin{align*}
    \cost(s) = \cost_f(s)+\cost_c(s),
\end{align*}
where $\cost_f(s)$ is the feedback cost, the total cost $\A'$ spends on buying signals when $s$ is drawn and $\cost_c(s)$ is the coverage cost, the number of actions taken by $\A'$ to cover $s$. That is to say
\begin{align*}
    \cost(\A',I) & = \E_{s \sim \D} \cost(s)= \E_{s \sim \D} \cost_f(s)+ \E_{s \sim \D}\cost_c(s) \\
    &\le 2 \E_{s \sim \D}\cost_c(s),
\end{align*}
since the feedback cost is always less than the coverage cost. 

In the rest of the proof, we will construct an instance $\Tilde{I}=(\X,\S,\ell,\D,\F')$ of stochastic optimization with time dependent feedback based on $I$ such that $\E_{s \sim \D}\cost_c(s) \le \alpha \opt(\Tilde{I})$ and $\opt(\Tilde{I}) \le \opt(I)$. We construct the feedback $\F'$ by constructing every possible sequence of signals received from $\F'$. Let $(y_0,y_1,\dots)$ be a sequence of signals received from the signaling schemes $(f_0,f_1,\dots)$. Let $c_t(y_t)$ be the cost to obtain signal $y_{t+1}$ for the sequence. Then for every $t \ge 0$, we make $c_t(y_t)$ copies for $y_t$. Thus, the corresponding signals sequence in $\F'$ is $(y_0,\dots,y_0,y_1,\dots,y_1,\dots)$, where $y_t$ appears $c_t(y_t)$ times.

Now we consider Algorithm~\ref{alg reduction}. If we ignore the step where we pay $c_t$ to get $y_{t+1}$, then the remaining algorithm is exactly running $\A$ over $\Tilde{I}=(\X,\S,\ell,\D,\F')$. Since $\A$ is an $\alpha$-competitive learner for $\Tilde{I}$, we know that 
\begin{align*}
    \E_{s \sim \D}\cost_c(s) = \cost(\A,\Tilde{I}) \le \alpha \opt(\Tilde{I}).
\end{align*}
It remains to show that $\opt(\Tilde{I}) \le \opt(I)$. Consider instance $I$. Assume $s$ is the drawn scenario and the corresponding sequence of signals is $(y_0,y_1,\dots)$. Assume the sequence of actions taken in $\opt(I)$ for this sequence of signals is $(a_0,a_1,\dots)$. We construct a sequence of actions taken for instance $\Tilde{I}$ with sequence of signals $(y_0,\dots,y_0,y_1,\dots,y_1,\dots)$ by modifying $(a_0,a_1,\dots)$. Assume that in $\opt(I)$, the learner pays a cost $c$ to obtain the next signal after taking actions $a_i$. Then in the modified sequence, we take $c$ arbitrary actions after $a_i$. For every drawn scenario $s$, the modified sequences can take less cost to cover $s$. This implies that $\opt(\Tilde{I}) \le \opt(I)$. Putting things together, we have 
\begin{align*}
    \cost(\A',I) \le 2 \E_{s \sim \D}\cost_c(s) \le 2\alpha \opt(\Tilde{I}) \le 2\alpha \opt(I), 
\end{align*}
which means Algorithm~\ref{alg reduction} is $2\alpha$-competitive for adaptive stochastic optimization with feedback.

\end{proof}


\subsection{Proof of Theorem~\ref{th time dependent up}}\label{sec pf greedy}

Before presenting the proof, we define the model of MSSC with time dependent feedback as a remainder.

\begin{definition}[Min Sum Set Cover with Time Dependent Feedback]
    Let $\B=[n]$ be a set of boxes, each box $i$ contains an unknown number $b_i \in \{0,1\}$
    A learner can know $b_i$ by querying box $i$, i.e. the action space $\X=\B$. A scenario $s \in \{0,1\}^n$ is a binary vector that represents the number contained in each box. If scenario $s$ is realized, then for every box $i \in \B$, $s_i=b_i$. A scenario $s$ is covered if a box $i$ such that $s_i=1$ is queried.
    Let $\S$ be a set of scenarios and $\D$ be a probability distribution over $\S$. Let $f$ be a sequence of feedback. A scenario $s^*$ is drawn from $\D$ initially. In each round $t$, a learner receives signal $y_t(s^*)$ from signaling scheme $f_t$ and takes an action $a_t \in \X$ to query the box $a_t$ and observed the number contained in that box. Given an instance $(\B,\S,\D,\F)$ of Min Sum Set Cover with Time Dependent Feedback, the goal of a learner is to construct the sequence of boxes $A$ to query to minimize $\E_{s\sim \D}\ell(A,s)$, where $\ell(A,s)$ is the number of boxes to query to cover the drawn scenario $s$.
\end{definition}

As a remainder, we restate Algorithm~\ref{alg greedy}, the simple greedy algorithm that we want to analyze here.
\begin{algorithm}[H]
		\caption{\textsc{Greedy} ($4$-competitive Learner for MSSC with Time Dependent Feedback)}
		\begin{algorithmic}
		
  \FOR{t=0,1,2,\dots}
  
\STATE Receive scenario set $S_t$ that are consistent with the feedback and outcomes received so far.
\STATE Compute $\Pr(i):=\sum_{s \in S_t:s_i=1}\Pr(s\mid S_t)$
\STATE Query any box $i^* \in \arg\max \{\Pr(i) \mid i \in \B\}$.
\IF {$s^*_{i^*}=1$}
\RETURN
\ENDIF

  \ENDFOR
		\end{algorithmic}
	\end{algorithm}

\begin{proof}

Without loss of generality, we can assume $\D$ is a uniform distribution over $\S$ and $\F$ contains only deterministic signaling schemes.
This is because given a distribution $\D$, we can modify $\S$ by making multiple copies of each scenario and uniformly draw a scenario from the modified set of scenarios according to our discussion in Appendix~\ref{sec signal}.
Under this assumption, we will write a linear program to lower bound $\opt(I)$. We say a scenario $s$ is covered by a box $i$ if $s_i=1$. For every scenario, $s$, denote by $L_s$ the set of boxes that cover $s$. 

Fix a sequence of feedback $\F$, let $T(\F)$ be the feedback tree induced by $\F$. Let $P(s)$ be the longest path in $T(\F)$ such that $s$ is contained in every node in $P(s)$.
Any learner $\A$ will assign a box to each node in $T(\F)$ such that for every scenario $s$, there is some node $v \in P(s)$ such that the box assigned to $v$ by $\A$ covers $s$. 
We derive the following integer program to capture the cost of a learner. For every node $v \in T(\F)$ and for every box $i \in \B$, let $x_{vi} \in \{0,1\}$ be the indicator if $\A$ assigns box $i$ to node $v$. For every node, $v \in T(\F)$ and for every scenario $s \in v$, let $y_{vs} \in \{0,1\}$ be the indicator if $s$ is not covered by any box assigned to an ancestor of $v$. Here, we use the notation $v'<v$ to denote that $v'$ is an ancestor of $v$.  For every scenario $s$, let $\cost(s):= \sum_{v \in P(s)}y_{vs}$, which is the time when $s$ is first covered by an assigned box. Then, any learner $\A$ gives a feasible solution to the following integer program.

\begin{align}
	\label{pr IP}
	\tag{IP}
	\begin{split}
		\min_{x,y} \ &  \sum_{s \in \S}\cost(s) \\
		\stt \ & \sum_{i \in \B}x_{vi} \le 1 \ \forall v \in T(\F) \\
		& y_{vs}+\sum_{i \in L_s}\sum_{v':v'<v}x_{v'i} \ge 1 \ \forall v \in T(\F), \forall s \in v \\
		& x_{vi} \in \{0,1\}\ \forall v\in T(\F), \forall i \in \B \\
  & y_{vs} \in \{0,1\} \  \forall v \in T(\F), \forall s \in \S.
	\end{split}
\end{align}
Here, the first set of constraints implies that for any node $v$, any learner can assign at most 1 box. The second set of constraints implies that for every node $v$ and every $s \in v$, either $s$ has not been covered so far or there is an ancestor $v'$ of $v$ that is assigned a box $i \in L_s$ by learner $\A$.
In particular, since $\D$ is uniform over $\S$, $\card{\S} \cost(\A,I) = \sum_{s \in \S}\cost(s)$. Thus, the following linear programming relaxation gives a natural lower bound for $\card{S}\opt(I)$.

\begin{align}
	\label{pr LP}
	\tag{LP}
	\begin{split}
		\min_{x,y} \ &  \sum_{s \in \S}\cost(s) \\
		\stt \ & \sum_{i \in \B}x_{vi} \le 1 \ \forall v \in T(\F) \\
		& y_{vs}+\sum_{i \in L_s}\sum_{v':v'<v}x_{v'i} \ge 1 \ \forall v \in T(\F), \forall s \in v \\
		& x_{vi} \ge 0\ \forall v\in T(\F), \forall i \in \B \\
  & y_{vs} \ge 0 \  \forall v \in T(\F), \forall s \in v.
	\end{split}
\end{align}
Let $\{A_v\}_{v \in T(\F)}$ be the set of dual variables for the first set of constraints in \eqref{pr LP} and let $\{B_{vs}\}_{v \in T(\F),s \in v}$ be the set of dual variables for the second set of constraints in \eqref{pr LP}. Then we derive the following dual linear program for \eqref{pr LP}.

\begin{align}
	\label{pr Dual}
	\tag{DUAL}
	\begin{split}
		\max_{A,B} \ &  \sum_{v \in T(\F)}\sum_{s \in v} B_{vs} - \sum_{v \in T(\F)}A_v \\
		\stt \ & B_{vs}\le 1 \ \forall v \in T(\F) \\
		& \sum_{s:s\in v, i \in L_s} \sum_{v':v<v'}B_{v's} \le A_v \ \forall v \in T(\F), \forall i \in \B \\
		& B_{vs} \ge 0\ \forall v\in T(\F), \forall s \in v \\
  & A_{v} \ge 0 \  \forall v \in T(\F).
	\end{split}
\end{align}
Since \eqref{pr LP} is feasible and bounded, we know from linear programming dual theory that \eqref{pr Dual} is feasible, furthermore, \eqref{pr Dual} and \eqref{pr LP} have the same optimal value. Denote by $D$ the optimal value of \eqref{pr Dual}, then we know that $\card{S}\opt(I) \ge D$.

Next, we will show that there is an optimal solution to \eqref{pr Dual} that has a special structure. We have the following observations.

\begin{observation}\label{obs structure1}
    Let $(A,B)$ be any feasible solution to \eqref{pr Dual}. For every $v \in T(\F)$, let $A'_v = \max_i \sum_{s: s\in v, i \in L_s}\sum_{v':v<v'}B_{v's}$. For every $v \in T(\F), s \in v$, let $B'_{vs}=B_{vs}$. Then $(A',B')$ is feasible to \eqref{pr Dual}, furthermore, $(A',B')$ has a larger objective value than $(A,B)$.
\end{observation}
The proof of Observation~\ref{obs structure1} follows directly by the second set of constraints in \eqref{pr Dual}.

\begin{observation}\label{obs structure2}
    Let $(A,B)$ be any feasible solution to \eqref{pr Dual}. For every $s \in \S$, let $C_s:=\sum_{v \in P(s)}B_{vs}$. For every $s\in \S$ and for every $v \in P(s)$, define 
    \begin{align*}
        B'_{vs} = \begin{cases}
            & 1 \ \text{ if } \dep(v) < \lfloor C_s\rfloor, \\
            & C_s - \lfloor C_s \rfloor \  \text{ if } \dep(v) = \lfloor C_s \rfloor, \\
            & 0 \ \text{ otherwise.}
        \end{cases}
    \end{align*}
    For every $v \in T(\F)$, define $A'_v = \max_i \sum_{s: s\in v, i \in L_s}\sum_{v':v<v'}B'_{v's}$. Then $(A',B')$ is feasible to \eqref{pr Dual} and $(A',B')$ has a larger objective value than $(A,B)$.
\end{observation}

\begin{opf}
    The feasibility of $(A',B')$ follows by Observation~\ref{obs structure1}. Thus, we only need to show $(A',B')$ has a larger objective value. We notice that
    \begin{align*}
        \sum_{v \in T(\F)}\sum_{s \in v} B_{vs} = \sum_{s \in \S}\sum_{v \in P(s)}B_{vs} = \sum_{s \in \S}C_s = \sum_{s \in \S}\sum_{v \in P(s)}B'_{vs}= \sum_{v \in T(\F)}\sum_{s \in v} B'_{vs}.
    \end{align*}
It remains to show that $\sum_{v \in T(\F)}A_v \ge \sum_{v \in T(\F)}A'_v.$ By Observation~\ref{obs structure1}, we may assume $A_v = \max_i \sum_{s: s\in v, i \in L_s}\sum_{v':v<v'}B_{v's}$ for every $v \in T(\F)$. It is sufficient to show for every $v$ and every $s \in v$, $\sum_{v' \in P(s):v<v'}B_{v's} \ge \sum_{v' \in P(s):v<v'}B'_{v's}$. We have 
\begin{align*}
    \sum_{v' \in P(s):v<v'}B_{v's} = C_s - \sum_{v' \in P(s):v'\le v}B_{v's} \ge C_s - \sum_{v' \in P(s):v'\le v}B_{v's} =  \sum_{v' \in P(s):v<v'}B'_{v's}.
\end{align*}
    
\end{opf}

Observation~\ref{obs structure1} and Observation~\ref{obs structure2} imply that an optimal solution to \eqref{pr Dual} can be constructed in the following way. For each $s\in \S$, assign $C_s \ge 0$ to $s$. For every $s\in \S$ and for every $v \in P(s)$, define 
    \begin{align*}
        B_{vs} = \begin{cases}
            & 1 \ \text{ if } \dep(v) < \lfloor C_s\rfloor, \\
            & C_s - \lfloor C_s \rfloor \  \text{ if } \dep(v) = \lfloor C_s \rfloor, \\
            & 0 \ \text{ otherwise.}
        \end{cases}
    \end{align*}
    For every $v \in T(\F)$, define $A_v = \max_i \sum_{s: s\in v, i \in L_s}\sum_{v':v<v'}B_{v's}$. Let $(A,B)$ be such a solution constructed in the way we discussed above using a vector $C \in R_+^\S$. Denote by $D(C)$ the objective value of $(A,B)$. Then we have
    \begin{align*}
        D(C) & = \sum_{v \in T(\F)}\sum_{s \in v} B_{vs} - \sum_{v \in T(\F)}\max_{i \in \B} \sum_{s: s\in v, i \in L_s}\sum_{v':v<v'}B_{v's} \\
        & = \sum_{s \in \S}C_s - \sum_{v \in T(\F)}\sum_{i \in \B} \sum_{s: s\in v, i \in L_s}\sum_{v':v<v'}B_{v's}\mathbf{1}_{vi} \\
        & = \sum_{s \in \S}C_s - \sum_{v \in T(\F)}\sum_{i \in \B} \sum_{v':v<v'} \sum_{s: s\in v, i \in L_s}B_{v's}\mathbf{1}_{vi} \\
        & = \sum_{s \in \S}C_s - \sum_{v \in T(\F)} \sum_{v':v<v'} \sum_{i \in \B} \sum_{s: s\in v, i \in L_s}B_{v's}\mathbf{1}_{vi} \\
        & = \sum_{s \in \S}C_s - \sum_{v' \in T(\F)} \sum_{v:v<v'} \sum_{i \in \B} \sum_{s: s\in v, i \in L_s}B_{v's}\mathbf{1}_{vi},
    \end{align*}
where $\mathbf{1}_{vi}$ is the indicator function if $i \in \arg \max_j \sum_{s: s\in v, j \in L_s}\sum_{v':v<v'}B_{v's}$. For convenience, we assume there is only one box that achieves the max.

We interpret $D(C)$ via the following physical process. For each scenario $s \in \S$, we generate a particle $\mathbf{P}_s$. $\mathbf{P}_s$ moves along the path $P(s)$ with a rate of $1$ and stops at time $t=C_s$. The length of an edge in $T(\F)$ is $1$. Let $V_s(t)$ be the speed of $\mathbf{P}_s$ at time $t$. That is to say $C_s = \int_{0}^\infty V_s(t) dt$ for every $s \in \S$.
From this point of view, we can write the first term in $D(C)$ as  
\begin{align}\label{eq C}
    \sum_{s \in \S} C_s = \int_{0}^\infty \sum_{s \in \S}V_s(t) dt.
\end{align}
On the other hand, for each node $v \in T(\F)$, there is a box $i(v)$ such that $\mathbf{1}_{vi(v)}=1$. At a given time $t$, we will charge each \textbf{moving} particle $G_s(t):=\card{\{v \mid v \in P(s), \dep(v) \le t, i(v) \in L_s\}}$. In other words, for every moving particle, we will charge it the number of visited nodes $v$ such that box $i(v)$ covers the corresponding scenario. Next, we build a connection between $G_s(t)$ and $D(C)$. For every $s \in \S$, write $P(s) = (v_0,v_1,\dots,v_n)$. Then
we have 
\begin{align*}
    G_s(t) = \sum_{i \in L_s}\sum_{v \in P(s): \dep(v) \le t}V_s(t)\mathbf{1}_{vi} = \sum_{i \in L_s}\sum_{j=0}^{\lfloor t\rfloor}V_s(t)\mathbf{1}_{v_ji}, 
\end{align*}
which implies 
\begin{align*}
    \int_0^\infty G_s(t) dt  = \int_0^\infty \sum_{i \in L_s}\sum_{j=0}^{\lfloor t\rfloor}V_s(t)\mathbf{1}_{v_ji}dt = \sum_{v' \in P(s)} \sum_{v: v \le v'} \sum_{i \in L_s}B_{v's}\mathbf{1}_{vi},
\end{align*}
according to the construction of $B$. Thus, we can write the second term in $D(C)$ as
\begin{align}\label{eq B}
    \sum_{v' \in T(\F)} \sum_{v:v<v'} \sum_{i \in \B} \sum_{s: s\in v, i \in L_s}B_{v's}\mathbf{1}_{vi} \le \int_0^\infty \sum_{s \in \S}G_s(t)dt.
\end{align}
Combine \eqref{eq C} and \eqref{eq B}, we get
\begin{align}\label{eq D}
    D(C) \ge \int_0^\infty \sum_{s\in \S} V_s(t) - \sum_{s \in \S}G_s(t) dt.
\end{align}

In the rest of the proof, instead of constructing the optimal solution to \eqref{pr Dual}, we will construct a vector $C_g$ based on Algorithm~\ref{alg greedy} such that $\card{S}\alg(I) \le 4D(C^g) \le 4D \le 4\card{S}\opt(I)$, which implies that Algorithm~\ref{alg greedy} is 4-competitive. 

Consider the implementation of Algorithm~\ref{alg greedy}, the greedy algorithm. We notice that if we arrive at some node $v \in T(\F)$, the set of scenarios $S_t$ we received is exactly $R_v$, the set of scenarios $s \in v$ that has not been covered so far. Since $\D$ is uniform, the box $\A(v)$ queried by the algorithm at node $v$ is the box that can cover most scenarios in $R_v$. 
Denote by $X_v=\{s \in R_v \mid \A(v) \in L_s\}$, which is the scenarios in $R_v$ covered by the box that Algorithm~\ref{alg greedy} queries in this round. Now we define $C_s$ for each scenario $s$. Let $P=(v_0,v_1,\dots,v_n)$ be a path of $T(\F)$ from the root to some leaf. For each $v_i \in P$, define $C_{v_i}=\max \{C_{v_{i-1}},\frac{\card{R_{v_i}}}{c\card{X_{v_{i}}}}\}$, where $c>0$ is a constant that we will determine later and $C_{v_0} = \frac{\card{R_{v_0}}}{c\card{X_{v_0}}}$.
Notice that $\{X_v\}_{v \in T(\F)}$ forms a partition of $\S$, so each $s$ belongs to a unique $X_v$. For every node $v \in T(\F)$ and for every $s \in v$, we set $C_s = C_v$. Denote by $C^g$ the vector we just constructed. We next show that $\card{S}\alg(I) \le 4D(C^g).$ 
Notice that 
\begin{align*}
    \card{\S}\alg(I) = \sum_{v\in T(\F)}(\dep(v)+1)\card{X_v} = \sum_{v\in T(\F)}\card{R_v}. 
\end{align*}
Based on this observation, we first derive the following lower bound for $\int_{0}^\infty \sum_{s \in \S}V_s(t) dt$. We have 
\begin{align*}
    \int_{0}^\infty \sum_{s \in \S}V_s(t) dt = \sum_{v \in T(\F)}\sum_{s \in X_v}\int_0^\infty V_s(s)dt=\sum_{v \in T(\F)}\sum_{s \in X_v}C_s \ge \frac{1}{c}\sum_{v \in T(\F)}\card{R_v}=\frac{1}{c}\card{\S}\alg(I).
\end{align*}

Next, we will show that $\int_0^\infty \sum_{s \in \S}G_s(t)dt \le \frac{1}{c}\int_{0}^\infty \sum_{s \in \S}V_s(t) dt.$ To do this, we upper bound $\sum_s G_s(t)$ for every $t \ge 0$. For every $s \in \S$, let $P^t(s)=(v_0(s),v_1(s),\dots,v_{\lceil t \rceil}(s))$ be the truncation of path $P(s)$ with length of $\lceil t \rceil$. We know that for every $t$, $P^t$, the set of such truncated paths, forms a partition of $\S$. So we can write $\sum_{s \in \S}G_s(t) = \sum_{P \in P^t}\sum_{s \in P}G_s(t)$. 

Let $P=(v_0,\dots,v_{\lceil t\rceil}) \in P^t$ be such a truncated path. The set of particles that are moving along $P$ corresponds to scenarios in $v_{\lceil t \rceil}$ with $C_s \ge t$. We observe that along the path $P$, $C_{v_i}$ is a step function with respect to the index $i$. Based on the definition of $C_s$, for every $v$ and every $s \in R_v$, we have $C_s \ge C_v$. This implies that along the path $P$, there must be some $i^* \le \lfloor t \rfloor$ such that the set of particles that are moving along $P$ at time $t$ corresponds to scenarios exactly in $R_{v_{i^*}} \cap v_{\lceil t\rceil}$. In particular, if we consider the set $P^t(i^*)$ of all paths in $P^t$ that passes $v_{i^*}$, then at time $t$, the set of particles moving along these paths is exactly $R_{v_{i^*}}$.

By the greedy property of Algorithm~\ref{alg greedy}, every box can cover at most $\card{X_{v_{i^*}}}$ scenarios from $R_{v_{i^*}} $. Since each path $P \in P^t(i^*)$ contains at most $t$ nodes and each node is charged by at most $\card{X_{v_{i^*}} }$ moving particles at time $t$, we have 
\begin{align*}
 \sum_{P \in P^t(i^*)}\sum_{s \in P}G_s(t) \le   t \card{X_{v_{i^*}}} \le  \frac{\card{R_{v_{i^*}}}}{c\card{X_{v_{i^*}}}} \card{X_{v_{i^*}}} = \frac{1}{c}\card{R_{v_{i^*}}} = \frac{1}{c}\sum_{P \in P^t(i^*)}\sum_{s \in P}V_s(t).
\end{align*}
Here, the second inequality follows by $C_{v_{i^*}} = \frac{\card{R_{v_{i^*}}}}{c\card{X_{v_{i^*}}}} \ge t$. The last equality holds because $\sum_{P \in P^t(i^*)}\sum_{s \in P}V_s(t)$ is the number of moving particles along paths in $P^t(i^*)$, which is $\card{R_{v_{i^*}}}$. Thus we have 
\begin{align*}
    \int_0^\infty \sum_{s \in \S}G_s(t)dt \le \frac{1}{c}\int_{0}^\infty \sum_{s \in \S}V_s(t) dt.
\end{align*}

Put the above discussions together, we have 
\begin{align*}
    D(C^g) \ge \int_0^\infty \sum_{s\in \S} V_s(t) - \sum_{s \in \S}G_s(t) dt \ge (1-\frac{1}{c})\int_{0}^\infty \sum_{s \in \S}V_s(t) dt \ge \frac{1}{c}(1-\frac{1}{c})\card{\S}\alg(I) = \frac{1}{4}\card{\S}\alg(I),
\end{align*}
by setting $c=2$ to maximize the ratio. This shows Algorithm~\ref{alg greedy} is $4$-competitive.

\end{proof}

\subsection{Proof of Theorem~\ref{th time dependent lb}}\label{sec pf time dependent lb}

\begin{proof}

We consider the following instance of min sum set cover with time dependent feedback. Let $\B$ be the set of $n$ boxes. The set of scenarios $\S = \{s^i\}_{i=1}^n$, where $s^i_j=1$ if $i=j$ and $0$ otherwise. $\D$ is a uniform distribution over $\S$. Let $\A$ be any deterministic learner. We design a set of feedback $\F_\A$ such that $\cost(\A,I) = \frac{n}{2}-o(1)$, while there is a learner $\A'$ such that $\cost(\A',I)=\frac{n}{4}+o(1)$. Here, $I=(\B,\S,\D,\F_\A)$ and $\cost(\A,I)$ is the cost of a learner $\A$ over instance $I$.

We describe $\F_A$ via its feedback tree representation $T(\F_\A)$. We first fix the structure of $T(\F_\A)$, then define the scenario contained in each node of $T(\F_\A)$. Let $T(\F_\A)$ be a binary tree. Let $v$ be a node in $T(\F_\A)$. We denote by $L(v)$ its left child and $R(v)$ its right child. Let $\{v_i\}_{i=1}^n$ be a path of $T(\F_\A)$ such that $v_{i+1}=R(v_i)$ and $v_1$ be the root of $T(\F_\A)$. We define the set of scenarios contained in each node in $T(\F)$. We know that $v_1 = \S$. Let $\A_i=\A(v_i)$ be the box queried by $\A$ at node $v_i$. We define $L(v_i)=\{s_{\A_i}\}$ and $R(v_i) = v_i \setminus L(v_i)$. This gives the definition of $\F_\A$.
Intuitively, every time $\A$ queries a box, $\F$ only tells $\A$ if it queries the unique box that contains $1$. This is to say $\F$ is useless for $\A$ and the cost of $\A$ is 
\begin{align*}
     \cost(\A,I) = \frac{1}{n} \sum_{j=1}^n j = \frac{n-1}{2}.
\end{align*}
On the other hand, let $\A'$ be the following learner. Let $\A'(v_i) = \A_{n+1-i}$, for $i \in [n]$ and $\A'(L(v_i))=\A_i$. That is, along the path $\{v_i\}_{i=1}^n$, the order of the queried box by $\A'$ is the inverse of that of $\A$ and at every node $L(v_i)$, $\A'$ queries the box corresponding to the unique scenario contained in $L(v_i)$. This implies 
\begin{align*}
    \cost(\A',I) = \frac{1}{n}+ \sum_{j=2}^{\frac{n-1}{2}} \frac{2j}{n} = \frac{n^2-5}{4n}.
\end{align*}
Thus, we have $\frac{ \cost(\A,I)}{ \cost(\A',I)} \to 2$, which implies no deterministic learner is $2-\epsilon$-competitive.

\end{proof}

\subsection{Proof of Theorem~\ref{th costly lb}}\label{sec pf costly lb}

Before presenting the proof, we remind the definition of buying information for MSSC.

\begin{definition}[Buying Information for Min Sum Set Cover]
    Let $(\B,\S,\D)$ be an instance of Min Sum Set Cover, $\F=\{f_t\}_{t=0}^\infty$ be a sequence of feedback and 
    $\C=\{c_t\}_{t=0}^\infty$ be a sequence of cost for receiving a signal from $f_{t+1}$ from $\F$. Initially, a scenario $s$ is drawn from $\D$. In each time round $t$, before $s$ is covered, a learner adaptively receives an arbitrary number of signals from the sequence $\F$ by paying the corresponding cost and then selects a box to query. An instance $(\B,\S,\D,\F,\C)$ of Buying Information for Min Sum Set Cover is to make decisions adaptively to minimize the expected number of the queried box plus the expected cost paid for the feedback to cover the random scenario.   
\end{definition}

\begin{proof}
    We consider the following instance of buying information for min sum set cover. Let $\B$ be the set of $n$ boxes. The set of scenarios $\S = \{s^i\}_{i=1}^n$, where $s^i_j=1$ if $i=j$ and $0$ otherwise. $\D$ is a uniform distribution over $\S$.  We assume the cost of obtaining any single feedback is $1$.
    Let $\A$ be any deterministic learner. We design a set of feedback $\F_\A$ for $\A$.

We describe $\F_A$ via its feedback tree representation $T(\F_\A)$. To do this, we will first fix the structure of $T(\F_\A)$, then describe the scenarios contained in each node. The structure of $T(\F_\A)$ is defined in the following way. There are $n_{i}+1$ nodes in $T(\F_\A)$ that have depth of $i$. Here $n_0 =0$ and for $i \ge 1$, $n_i \ge 0$ is a number that depends on $\A$. Furthermore, for each level of $T(\F_\A)$, only the rightmost node has children. In particular, for $i \ge 1$, let $v_{i+1}=R(v_i)$ be the right most child of $v_i$, where $v_1$ is the root of $T(\F_\A)$.

We notice that given the structure of $T(\F_\A)$, any deterministic learner $\A$ can be described in the following way using $T(\F_\A)$. For every node $v \in T(\F_\A)$, $\A$ will query a set of boxes $\B^v$ in some order, where $\card{\B^v} \ge 0$. Denote by $\B^i$, the set of boxes queried by $\A$ at node $v_i$. Let $n_i=\card{\B^i} \ge 0$, then set of scenarios that contained in $R(v_i)$ is defined by $\{s^j \in v_i \mid j \not\in \B^i\}$.
Recall that there are $n_i+1$ nodes in $T(\F_\A)$ that have depth $i$ and we have defined the set of scenarios contained in one of these nodes. For the rest of $n_i$ nodes, we assign a unique scenario covered by $\B^i$ to each of them. This gives the definition of $\F_\A$.
In particular, $\F_\A$ is useless for $\A$, since every time $\A$ asks for feedback, the feedback only tells $\A$ which scenarios are not covered so far.

Now we compute the cost of $\A$. Consider the path $(v_1,v_2,\dots,v_k)$ in $T(\F_\A)$, such that $\sum_{i=1}^k\card{\B^i}=n$. That is to say, all scenarios are covered before the $k$th feedback is asked. Notice that $(\B^1,\dots,\B^k)$ forms a partition of $\S$. Let $s \in \B^i$ be the $j$th scenario in $\B^i$ covered by $\A$, then the cost of $\A$ when scenario $s$ is drawn is 
\begin{align*}
    \cost(\A,s)=\sum_{\ell=1}^{i-1}n_\ell+i-1+j,
\end{align*}
which implies
\begin{align*}
    \cost(\A,I) = \E_s \cost(\A,s) &= \frac{1}{n}\sum_{i=1}^k\sum_{j=1}^{n_i}\left(\sum_{\ell=1}^{i-1}n_\ell+i-1+j\right) \\
    & = \frac{1}{n} \left(\sum_{i=1}^n i + \sum_{i=1}^k\sum_{j=1}^{n_i} i-n\right).
\end{align*}
We consider the two different cases. In the first case, $\sum_{i=1}^n i \le \sum_{i=1}^k\sum_{j=1}^{n_i} i$. We notice that any deterministic learner $\A^*$ that asks for no feedback has a cost $\E_s\cost(\A^*,s)=\frac{1}{n}\sum_{i=1}^n i$. This means
\begin{align*}
    \frac{ \cost(\A,I)}{\cost(\A^*,I)} \ge \frac{\frac{2}{n}\sum_{i=1}^n i-1}{\frac{1}{n}\sum_{i=1}^n i}=2-o_n(1).
\end{align*}
In the second case, we assume $\sum_{i=1}^n i > \sum_{i=1}^k\sum_{j=1}^{n_i} i$. In this case, we define a deterministic learner $\A^*$ in the following way. $\A^*$ keeps asking for feedback until the feedback reveals the drawn scenario, then $\A^*$ covers the drawn scenario via the unique box. It is not hard to see, any scenario in $\B^i$ will cost $\A^*$, $i+1$. Thus, $\E_s\cost(\A^*,s) =1+ \frac{1}{n}\sum_{i=1}^k\sum_{j=1}^{n_i} i$. In this case, we have 
\begin{align*}
    \frac{ \cost(\A,I)}{\cost(\A^*,I)} \ge \frac{\frac{1}{n} \left(\sum_{i=1}^n i + \sum_{i=1}^k\sum_{j=1}^{n_i} i\right)-1}{1+ \frac{1}{n}\sum_{i=1}^k\sum_{j=1}^{n_i} i}= \frac{\frac{1}{n} \left(\sum_{i=1}^n i + \sum_{i=1}^k\sum_{j=1}^{n_i} i\right)}{\frac{1}{n}\sum_{i=1}^k\sum_{j=1}^{n_i} i}-o_n(1) \ge 2-o_n(1).
\end{align*}

Thus, for every $\epsilon>0$, there is no deterministic learner that is $2-\epsilon$ competitive.

\end{proof}


\end{document}